\documentclass{article}

\usepackage{amssymb}     
\usepackage{marvosym}    
\usepackage{upgreek}     
\newcommand{\ctsper}      {cts/(keV$\cdot$kg$\cdot$yr)}

\newcommand{\kgy}         {{kg$\cdot$yr}}
\newcommand{\kgyr}        {{kg$\cdot$yr}}

\newcommand{\mum}         {{$\upmu$m}}

\newcommand{\qbb}         {{$Q_{\beta\beta}$}}

\newcommand{\thalftwo}    {${T^{2\nu}_{1/2}}$}

\newcommand{\onbb}        {{$0\nu\beta\beta$}}

\newcommand{\nnbb}        {{$2\nu\beta\beta$}}

\newcommand{\twonu}       {{$2\nu\beta\beta$}}

\newcommand{\gerda}       {\textsc{Gerda}}

\newcommand{\GERDA}       {\mbox{\textsc{Gerda}}}

\newcommand{\phaseone}    {Phase~I}
\newcommand{\phasetwo}    {Phase~II}

\newcommand{\gerdatwo}    {\gerda\ Phase~II}

\newcommand{\igex}        {\textsc{Igex}}

\newcommand{\hdm}         {\textsc{HdM}}

\newcommand{\geant}       {\textsc{Geant4}}

\newcommand{\mage}        {\textsc{MaGe}}

\newcommand{\gesix}       {{$^{76}$Ge}}

\newcommand{\radzzs}      {{$^{226}$Ra}}

\newcommand{\kvn}         {{$^{40}$K}}
\newcommand{\kvz}         {{$^{42}$K}}
\newcommand{\Am}          {$^{241}$Am}
\newcommand{\Rn}          {$^{222}$Rn}
\newcommand{\Ra}          {$^{226}$Ra}
\newcommand{\Po}          {$^{210}$Po}
\newcommand{\Ar}          {$^{39}$Ar}
\newcommand{\Kr}          {$^{85}$Kr}

\newcommand{\Bil}         {$^{212}$Bi}
\newcommand{\Bih}         {$^{214}$Bi}

\newcommand{\Th}          {$^{228}$Th}
\newcommand{\Tl}          {$^{208}$Tl}

\newcommand{\Uh}          {$^{238}$U}

\newcommand{\Co}          {$^{60}$Co}
\newcommand{\Ac}          {$^{228}$Ac}
\newcommand{\Pbl}         {$^{210}$Pb}
\newcommand{\Pbh}         {$^{214}$Pb}
\newcommand{\Pa}          {$^{234\text{m}}$Pa}

\pdfoutput=1 

\usepackage[utf8]{inputenc}
\usepackage[T1]{fontenc}
\usepackage{dcolumn}
\usepackage[dvipsnames]{xcolor}
\usepackage{graphicx}
\usepackage{subfig}
\usepackage{hyperref}
\usepackage[mode=buildnew]{standalone}
\usepackage{upgreek}
\usepackage{booktabs}
\usepackage{colortbl}
\usepackage{longtable}
\usepackage{rotating}
\usepackage{multirow, makecell}
\usepackage{verbatim}
\usepackage[parfill]{parskip}
\usepackage{tikz}
\usepackage{siunitx}
\usepackage{capt-of}
\usepackage{pdflscape}
\usepackage{afterpage}
\usepackage{array}

\usepackage[margin=2cm]{geometry} 

\usepackage{amsmath}
\renewcommand{\epsilon}{\varepsilon}
\renewcommand{\theta}{\vartheta}
\renewcommand{\rho}{\varrho}
\renewcommand{\phi}{\varphi}
\usepackage{mathtools}

\graphicspath{%
  {./0-intr/img/}%
  {./1-glob/img/}%
  {./2-resu/img/}%
  {./3-disc/img/}%
  {./4-conc/img/}%
  {./apdx/img/}%
}

\newcommand{\pplus}{$\text{p}^+$}
\newcommand{\nplus}{$\text{n}^+$}

\newcommand{\enrBEGe}{\texttt{M1-enrBEGe}}
\newcommand{\enrCoax}{\texttt{M1-enrCoax}}
\newcommand{\enrGe}{\texttt{M2-enrGe}}
\newcommand{\pvalue}{\textit{p}-value}

\newcommand{\m}[1]{\texttt{#1}}

\newcommand{\Mokvn}{\m{M1-K40}}
\newcommand{\Mtkvn}{\m{M2-K40}}
\newcommand{\Mokvz}{\m{M1-K42}}
\newcommand{\Mtkvz}{\m{M2-K42}}

\newcommand{\tetratex}{Tetratex\textsuperscript{\textregistered}}

\newcommand{\Thc}{$^{232}$Th}

\begin{document}

\title{Modeling of \gerda\ \phasetwo\ data}
\date{\vspace{-5ex}}

\maketitle
\thispagestyle{empty}

\newcommand\ALNGS{\,{\bf $^{a}$}}
\newcommand\AQU{\,{\bf $^{b}$}}
\newcommand\CAT{\,{\bf $^{c}$}}
\newcommand\CR{\,{\bf $^{d}$}}
\newcommand\DD{\,{\bf $^{e}$}}
\newcommand\JINR{\,{\bf $^{f}$}}
\newcommand\GEEL{\,{\bf $^{g}$}}
\newcommand\HD{\,{\bf $^{h}$}}
\newcommand\MIBF{\,{\bf $^{i}$}}
\newcommand\MIBINFN{\,{\bf $^{j}$}}
\newcommand\MILUINFN{\,{\bf $^{k}$}}
\newcommand\INR{\,{\bf $^{l}$}}
\newcommand\ITEP{\,{\bf $^{m}$}}
\newcommand\KU{\,{\bf $^{n}$}}
\newcommand\ITEPKU{\,{\bf $^{m,n}$}}
\newcommand\MPIP{\,{\bf $^{o}$}}
\newcommand\TUM{\,{\bf $^{p}$}}
\newcommand\PDUNI{\,{\bf $^{q}$}}
\newcommand\PDINFN{\,{\bf $^{r}$}}
\newcommand\TU{\,{\bf $^{s}$}}
\newcommand\UZH{\,{\bf $^{t}$}}
\newcommand\JINRINR{\,{\bf $^{f,l}$}}
\newcommand\ITEPINR{\,{\bf $^{m,l}$}}
\newcommand\CRJINR{\,{\bf $^{d,f}$}}
\newcommand\PDUNINFN{\,{\bf $^{q,r}$}}
\newcommand\MIBFINFN{\,{\bf $^{i,j}$}}
\newcommand\KUJINR{\,{\bf $^{n,f}$}}
\newcommand\KUJINRTUM{\,{\bf $^{n,f,p}$}}
\newcommand\JINRKUTUM{\,{\bf $^{f,n,p}$}}
\newcommand\HDLNGS{\,{\bf $^{h,a}$}}
\newcommand\HDJINR{\,{\bf $^{h,f}$}}
\newcommand\INRHD{\,{\bf $^{l,h}$}}
\newcommand\LNGSMIL{\,{\bf $^{a,i}$}}
\newcommand\HDH{\,{\bf $^{h}$}}
\newcommand\HDL{\,{\bf $^{h}$}}
\newcommand\SJU{\,{\bf $^{`}$}}
\newcommand\KO{\,{\bf $^{}$}}

\newcommand\lngsasse{{$^a$)}\,}
\newcommand\aqu{{$^b$)}\,}
\newcommand\cat{{$^c$)}\,}
\newcommand\cra{{$^d$)}\,}
\newcommand\dd{{$^e$)}\,}
\newcommand\jinr{{$^f$)}\,}
\newcommand\geel{{$^g$)}\,}
\newcommand\hd{{$^h$)}\,}
\newcommand\mibf{{$^i$)}\,}
\newcommand\mibinfn{{$^j$)}\,}
\newcommand\miluinfn{{$^k$)}\,}
\newcommand\inr{{$^l$)}\,}
\newcommand\itep{{$^m$)}\,}
\newcommand\ku{{$^n$)}\,}
\newcommand\mpipmun{{$^o$)}\,}
\newcommand\tum{{$^p$)}\,}
\newcommand\pduni{{$^q$)}\,}
\newcommand\pdinfn{{$^r$)}\,}
\newcommand\tu{{$^s$)}\,}
\newcommand\uzh{{$^t$)}\,}

\newcommand{\upind}[1]{$^{#1}$\,}
\makeatletter
\def\note{\xdef\@thefnmark{}\@footnotetext}
\makeatother

\begin{center}
  \noindent
  Gerda~collaboration\upind{1}\note{\upind{1} \emph{email:} gerda-eb{@}mpi-hd.mpg.de}{: }
  M.~Agostini\TUM{, }
  A.M.~Bakalyarov\KU{, }
  M.~Balata\ALNGS{, }
  I.~Barabanov\INR{, }
  L.~Baudis\UZH{, }
  C.~Bauer\HD{, }
  E.~Bellotti\MIBFINFN{, }
  S.~Belogurov\ITEPINR\upind{,2}\note{\upind{2} \emph{also at:} NRNU MEPhI, Moscow, Russia}{, }
  A.~Bettini\PDUNINFN{, }
  L.~Bezrukov\INR{, }
% ...
\end{center}
\vspace{11pt}
\begin{center}
  {\lngsasse INFN Laboratori Nazionali del Gran Sasso and Gran Sasso Science Institute, Assergi, Italy}             \\[1mm]
  {\aqu      INFN Laboratori Nazionali del Gran Sasso and Universit{\`a} degli Studi dell'Aquila, L'Aquila, Italy}  \\[1mm]
  {\cat      INFN Laboratori Nazionali del Sud, Catania, Italy}                                                     \\[1mm]
  {\cra      Institute of Physics, Jagiellonian University, Cracow, Poland}                                         \\[1mm]
  {\dd       Institut f{\"u}r Kern- und Teilchenphysik, Technische Universit{\"a}t Dresden, Dresden, Germany}       \\[1mm]
  {\jinr     Joint Institute for Nuclear Research, Dubna, Russia}                                                   \\[1mm]
  {\geel     European Commission, JRC-Geel, Geel, Belgium}                                                          \\[1mm]
  {\hd       Max-Planck-Institut f{\"u}r Kernphysik, Heidelberg, Germany}                                           \\[1mm]
  {\mibf     Dipartimento di Fisica, Universit{\`a} Milano Bicocca, Milan, Italy}                                   \\[1mm]
  {\mibinfn  INFN Milano Bicocca, Milan, Italy}                                                                     \\[1mm]
  {\miluinfn Dipartimento di Fisica, Universit{\`a} degli Studi di Milano and INFN Milano, Milan, Italy}            \\[1mm]
  {\inr      Institute for Nuclear Research of the Russian Academy of Sciences, Moscow, Russia}                     \\[1mm]
  {\itep     Institute for Theoretical and Experimental Physics, NRC ``Kurchatov Institute'', Moscow, Russia}       \\[1mm]
  {\ku       National Research Centre ``Kurchatov Institute'', Moscow, Russia}                                      \\[1mm]
  {\mpipmun  Max-Planck-Institut f{\"ur} Physik, Munich, Germany}                                                   \\[1mm]
  {\tum      Physik Department, Technische  Universit{\"a}t M{\"u}nchen, Germany}   \\[1mm]
  {\pduni    Dipartimento di Fisica e Astronomia, Universit{\`a} degli Studi di Padova, Padua, Italy}               \\[1mm]
  {\pdinfn   INFN Padova, Padua, Italy}                                                                             \\[1mm]
  {\tu       Physikalisches Institut, Eberhard Karls Universit{\"a}t T{\"u}bingen, T{\"u}bingen, Germany}           \\[1mm]
  {\uzh      Physik-Institut, Universit{\"a}t Z{\"u}rich, Z{u}rich, Switzerland}                                    \\[1mm]
\end{center}

\cleardoublepage
\setcounter{page}{1}

\begin{abstract}
  The GERmanium Detector Array (\GERDA) experiment at the Gran Sasso
  underground laboratory (LNGS) of INFN is searching for neutrinoless
  double-beta (\onbb) decay of \gesix. The technological challenge of
  \gerda\ is to operate in a ``background-free'' regime in the region of
  interest (ROI) after analysis cuts for the full 100~\kgy\ target
  exposure of the experiment. A careful modeling and decomposition of
  the full-range energy spectrum is essential to predict the shape and
  composition of events in the ROI around \qbb\ for the \onbb\ search,
  to extract a precise measurement of the half-life of the double-beta
  decay mode with neutrinos (\nnbb) and in order to identify the
  location of residual impurities. The latter will permit future
  experiments to build strategies in order to further lower the
  background and achieve even better sensitivities. In this article the
  background decomposition prior to analysis cuts is presented for
  \gerdatwo.  The background model fit yields a flat spectrum in the ROI
  with a background index (BI) of $16.04^{+0.78}_{-0.85} \cdot
  10^{-3}$~\ctsper\ for the enriched BEGe data set and
  $14.68^{+0.47}_{-0.52} \cdot 10^{-3}$~\ctsper\ for the enriched
  coaxial data set. These values are similar to the one of \phaseone\
  despite a much larger number of detectors and hence radioactive
  hardware components.
\end{abstract}


\section{Introduction}%
\label{sec:intro}

A large fraction of current experimental efforts are devoted to test the
precision of the standard model of particle physics and investigate the
presence of new phenomena. Many extensions of the standard model predict
rare processes and in particular the existence of neutrinoless
double-beta (\onbb) decay~\cite{mohapatra2006,mohapatra2007,pas2015}.
The observation of this lepton-number violating decay would shed light
on the nature of neutrinos and could give a hint on the scale of
neutrino masses.

The GERmanium Detector Array (\gerda)
experiment~\cite{Ackermann2013,Agostini2018a} is searching for \onbb\
decay of the candidate isotope \gesix\ at a Q-value of $Q_{\beta\beta} =
2039.061(7)$~keV~\cite{Mount2010}. \gerda\ is operating 37 detectors
made from material enriched in \gesix\ and a total mass of 35.6~kg bare
in 64~m$^3$ of liquid Argon (LAr, purity 5.0). The experiment profits
from the high shielding power of the LAr and its scintillation
properties. A hybrid instrumentation consisting of light guiding fibers
read out by silicon photomultipliers (SiPMs) and 16 photomultipliers
(PMTs) detect LAr scintillation light in order to veto events
depositing energy in the cryogenic liquid~\cite{Agostini2018a}. The LAr
cryostat itself is situated inside a tank filled with 590~m$^3$ of
purified water shielding against external ionizing radiation and
neutrons. Furthermore, it is instrumented with 66 PMTs to veto muons by
the detection of \v{C}erenkov light. \gerda\ is the first \onbb\ decay
experiment working in a ``background-free'' regime in the region of
interest (ROI) after analysis cuts~\cite{Agostini2017, Agostini2018,
Agostini2019}, where ROI is $Q_{\beta\beta} \pm \text{FWHM}/2$, and FWHM
is full width half maximum.

In the following, we present the spectral decomposition of \gerdatwo\
data. The analysis is conducted prior the application of active
background suppression techniques to data, i.e.~the LAr
veto~\cite{Agostini2018a} and pulse shape discrimination (PSD) taking
advantage of particular detector signal shapes~\cite{Agostini2013}. A
new assay of the \gerda\ background is necessary due to substantial
upgrade works finished in 2015~\cite{Agostini2018a}. Most structural
components close to the detectors have been exchanged using materials
with improved radio-purity, the detector array has been enlarged and the
LAr veto instrumentation has been deployed during the upgrade. Moreover,
each detector string (enclosed in a copper mini-shroud during \phaseone)
has been encapsulated in a transparent nylon mini-shroud in order to
limit the drift of \kvz\ ions in the detector vicinity and appropriately
propagate the LAr scintillation light~\cite{Lubashevskiy2018} (see
\autoref{subsec:priors} for details). The introduction of these new
setup components and materials changes the distribution and composition
of radioactive impurities in the setup.

A precise knowledge of the spectral composition of the data is a key
point for further analysis like accessing the half-life of the lepton
number conserving mode of double-beta (\twonu) decay. Moreover, there
are significant efforts towards reaching the tonne-scale of active
isotope mass and the localization of remaining radioactive impurities
inside the setup is the basis for the possible further reduction of
background. This is essential for future endeavors in order to boost the
current signal discovery and limit setting sensitivity by two orders of
magnitude to the range of $T^{0\nu}_{1/2} > 1\cdot10^{28}$~yr.


\section{Data selection and prior knowledge}%
\label{sec:global}\label{subsec:data}

The data analyzed in the following were taken between December 2015 and
April 2018. In this period the \gerda\ array consisted of 40 high-purity
germanium (HPGe) detectors: 30 Broad Energy Germanium (BEGe)
detectors~\cite{Agostini2015e,GERDAcollaboration2019} and 10 detectors
with a semi-coaxial geometry three of which are made from germanium
with a natural isotope composition. The enrichment fraction of the 30
enriched BEGe ($^\text{enr}$BEGe) detectors is $87.8$\% while the
respective fraction for the 7 enriched coaxial ($^\text{enr}$Coax)
detectors is in the range of $85.5 - 88.3$ \%~\cite{Agostini2018a}.

\subsection{Detector geometries}

The \gerda\ HPGe detectors are made of p-type germanium. \pplus\ and
\nplus\ contacts are manufactured via boron implantation and lithium
diffusion, respectively. The \pplus\ electrode is connected to a charge
sensitive amplifier while the \nplus\ electrode is biased at typically
4~kV. A groove between the two contacts provides electrical insulation.
The bias high-voltage creates an internal electrical field which is
responsible for charge collection. When biased at full-depletion
voltage, the germanium detectors reach maximal ($\epsilon = 1$) charge
collection efficiency (CCE) in an internal active volume, surrounded by
a transition layer (TL) with reduced CCE ($0 < \epsilon < 1$) and low
electric field. The TL is covered by a thin conductive layer in which
all charges recombine and charge collection is entirely suppressed
($\epsilon = 0$), therefore, called dead layer. We define the contact
thickness as the depth at which the CCE reaches its maximal value. The
\gerda\ detectors are of two distinct geometries. In the semi-coaxial
layout the thin \pplus\ contact ($0.5-1~\upmu$m) covers the entire bore
hole; in the BEGe-type, instead, the same contact is a disk of 15~mm
diameter (see figure~3 in reference~\cite{Agostini2014}). The \nplus\
contact, about 1~mm thick, ``wraps around'' the detector. An exhaustive
description of the \gerda\ detector geometries and properties can be
found in previous
publications~\cite{Agostini2018a,Agostini2014,Agostini2015e,GERDAcollaboration2019}.
The detector arrangement in the 7 strings that constitute the \gerda\
array is graphically presented in~\autoref{fig:magevolumes}a (and in the
appendix in~\autoref{fig:detstrings}).

\subsection{Data acquisition and treatment}

All data are recorded using FADCs and are digitally processed
off-line~\cite{Agostini2018a}. The linearity of the data acquisition
system and off-line energy reconstruction was tested with a precision
pulse generator over the whole dynamic range of the FADCs. Up to an
energy of at least 6~MeV no major non-linearity and pulse shape
deformation was observed. A signal above a threshold of about $100$~keV
in any of the germanium detectors triggers the data acquisition and the
respective event is written to disk.\footnote{The exact threshold is
detector and run dependent and varies between 20~keV and
200~keV~\cite{vanhoefen2018}.} An event is defined as the set of traces
recorded in the 40 germanium detectors, 16 photomultipliers (PMT) and 15
silicon photomultiplier (SiPM) channels from the LAr veto and the signal
from the Water \v{C}erenkov muon veto. In the following, we define the
multiplicity of an event as the number of germanium detectors in which
an energy of at least 40~keV is registered.\footnote{Note that this
energy threshold is lowered to 5~keV in the \onbb\ analysis.}

The energy deposition associated to each germanium detector signal is
determined via a zero area cusp (ZAC) filter~\cite{Agostini2015} which
is optimized off-line for each detector and each calibration.
Calibrations are usually taken with three \Th\ sources which are lowered
into the LAr to the vicinity of the detector array in a 1--2 week
cycle. An energy correction due to crosstalk between detector channels
is performed for each event. The average crosstalk for all pairs of
channels is about 0.05\%. Details about the crosstalk correction can be
found in reference~\cite{wester2019}. Events in a window $Q_{\beta\beta}
\pm 25$~keV are excluded from the analysis until all selection cuts are
finalized. The number of events and their energies in this window are
only released once all analysis steps are defined.

Each event has to pass a number of quality cuts which are tailored to
filter unphysical events~\cite{Agostini2017}. Data taking periods in
which stable operation cannot be guaranteed are excluded from analysis.
Detectors with an unstable energy calibration are used only to determine
the event multiplicity but do not enter any data set, e.g.~an event that
triggers three detectors one of which cannot be calibrated well is not
considered a two- but a three-detector event. Also, two-detector events
involving a detector which is not well calibrated are rejected. Events
with a multiplicity higher than two are discarded by default and,
likewise, events which trigger the muon veto are excluded. 

\subsection{Analysis data sets}

\begin{table}[tb]
  \centering
  \caption{%
    Properties of the data sets considered in this analysis. Further
    details about the \gerda\ detectors can be found in past
    publications~\cite{Agostini2014,GERDAcollaboration2019}.%
  }\label{tab:datasetdesc}
  \begin{tabular}{lccccc}
  \toprule
  \multirow{2}{*}{data set}  & \multirow{2}{*}{composition}      & total Ge           & active \gesix\   & total Ge           & active \gesix\   \\
                             &                                   & mass [kg]          & mass [kg]        & exposure [\kgyr]   & exposure [\kgyr] \\
  \midrule
  \enrBEGe\                  & 29 $^\text{enr}$BEGe\footnotemark & $19.362 \pm 0.029$ & $15.06 \pm 0.40$ & $32.124 \pm 0.048$ & $25.08 \pm 0.45$ \\
  \enrCoax\                  & 7 $^\text{enr}$Coax               & $15.576 \pm 0.007$ & $11.61 \pm 0.54$ & $28.088 \pm 0.013$ & $21.0  \pm 1.0$  \\
  \enrGe\                    & all enriched                      & $34.938 \pm 0.030$ & $26.67 \pm 0.67$ & $60.212 \pm 0.050$ & $46.1  \pm 1.1$  \\
  \bottomrule
\end{tabular}


\end{table}%
\footnotetext[3]{The BEGe detector \texttt{GD02D} is the only detector
  that does not fully deplete~\cite{GERDAcollaboration2019}. Hence,
  events triggered by this detector are not considered in either data
  set and it is omitted from the mass computation.}

Events of multiplicity one (\texttt{M1}) and multiplicity two
(\texttt{M2}) from detectors with enriched isotope composition are
accounted for in the construction of the analysis data sets. Events from
the coaxial detectors with natural isotope composition, located in the
central detector string, are not used in this analysis due to large
uncertainties on their n$^+$ contact thickness and detection efficiency.
The \texttt{M1} events are split in two data sets based on the two
enriched detector geometries which we call \enrBEGe\ and \enrCoax\ in
the following. The \texttt{M2} data form a third data set which is named
\enrGe. The energy we associate to an \texttt{M2} event is the sum of
the energies reconstructed in the two detectors. The data sets, their
exposure and respective detector mass are listed
in~\autoref{tab:datasetdesc}.

\subsection{Monte Carlo simulations and probability density functions}

\begin{figure}[tb]
  \centering
  \includegraphics{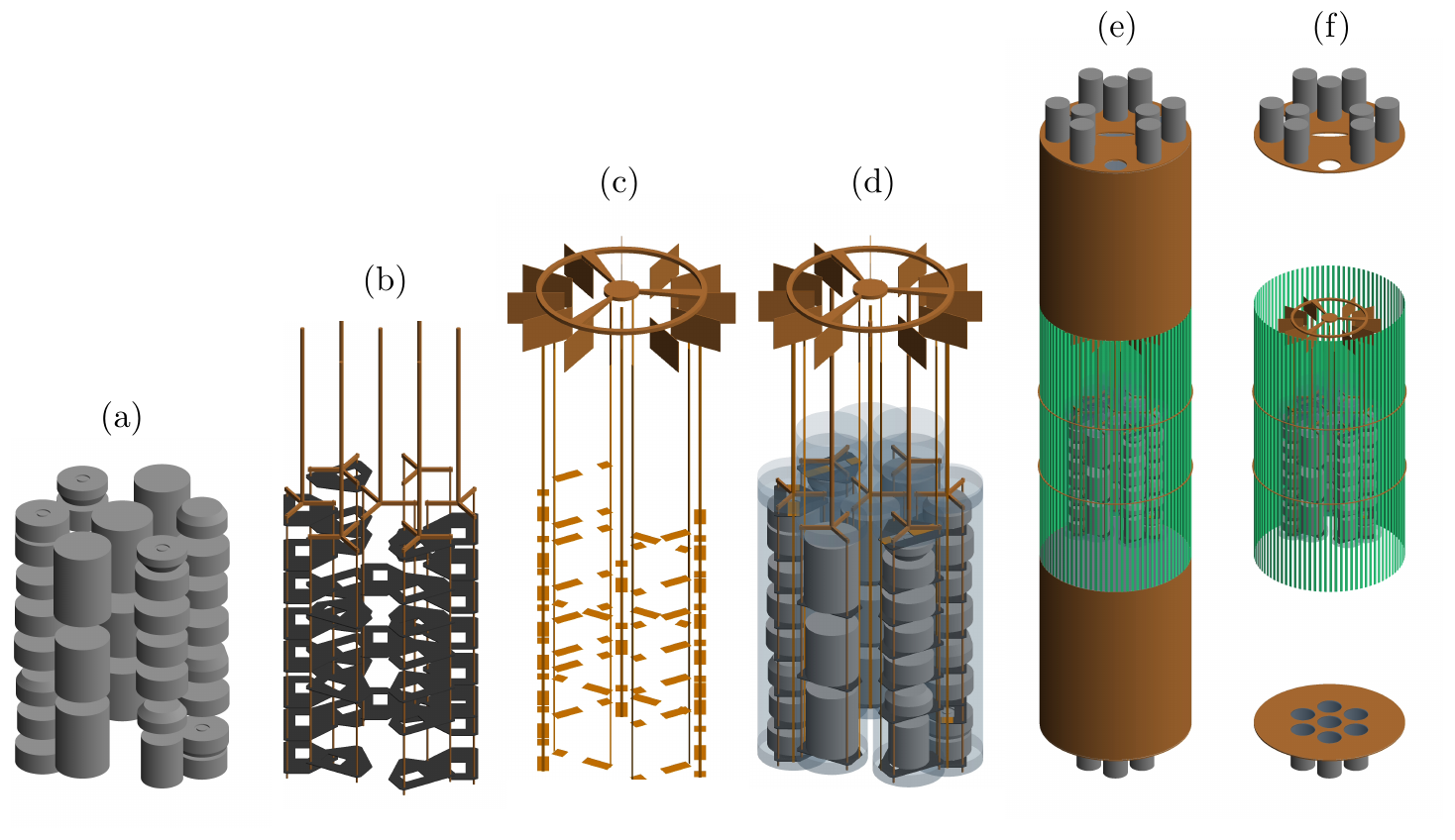}
  \caption{%
    Implementation of the \gerda\ array in \mage, visualized using the
    \geant\ visualization drivers. From left to right: a) the \gerda\
    detectors, b) the holder mounting, composed of silicon plates and
    copper bars c) the high-voltage and signal flexible flat cables plus
    the front-end electronics on top, d) the full array instrumentation,
    including the transparent nylon mini-shrouds, e) the full LAr veto
    system surrounding the array, including the fiber shroud (in green),
    the \tetratex-coated copper shrouds (above and below the fibers) and
    the two PMT arrays, f) the LAr veto system without the copper
    shrouds.%
  }\label{fig:magevolumes}
\end{figure}

The Probability Density Functions (PDFs) used to model contributions to
the energy spectra are obtained from Monte Carlo simulations. The
latter are performed using the \mage\ simulation
framework~\cite{boswell2011}, based on
\geant~\texttt{v10.4}~\cite{agostinelli2002,allison2006,Allison2016}.
\mage\ contains a software implementation of the \gerdatwo\ detectors as
well as the assembly and all other surrounding hardware components. A
visualization of this implementation is presented in
\autoref{fig:magevolumes}. Detector intrinsic \twonu\ decays of \gesix\
and background events originating from radioactive contaminations in and
around the detector assembly are simulated. The energy spectrum of the
two electrons emitted in the \twonu\ decay was sampled according to the
distribution given in reference~\cite{Tretyak1995} implemented in
\textsc{Decay0}~\cite{Ponkratenko2000}. The PDFs are obtained from the
Monte Carlo simulations, taking into account the finite energy
resolution and individual exposure acquired with each detector during
the considered data taking periods. Special care is taken not to
statistically bias the PDFs by assuring that each simulated decay is
taken into account only once in the production of a PDF. For more
details see \autoref{apdx:pdfs}.



\subsection{Background expectation}%
\label{subsec:priors}

\begin{figure}[tp]
  \centering
  \includegraphics[width=\textwidth]{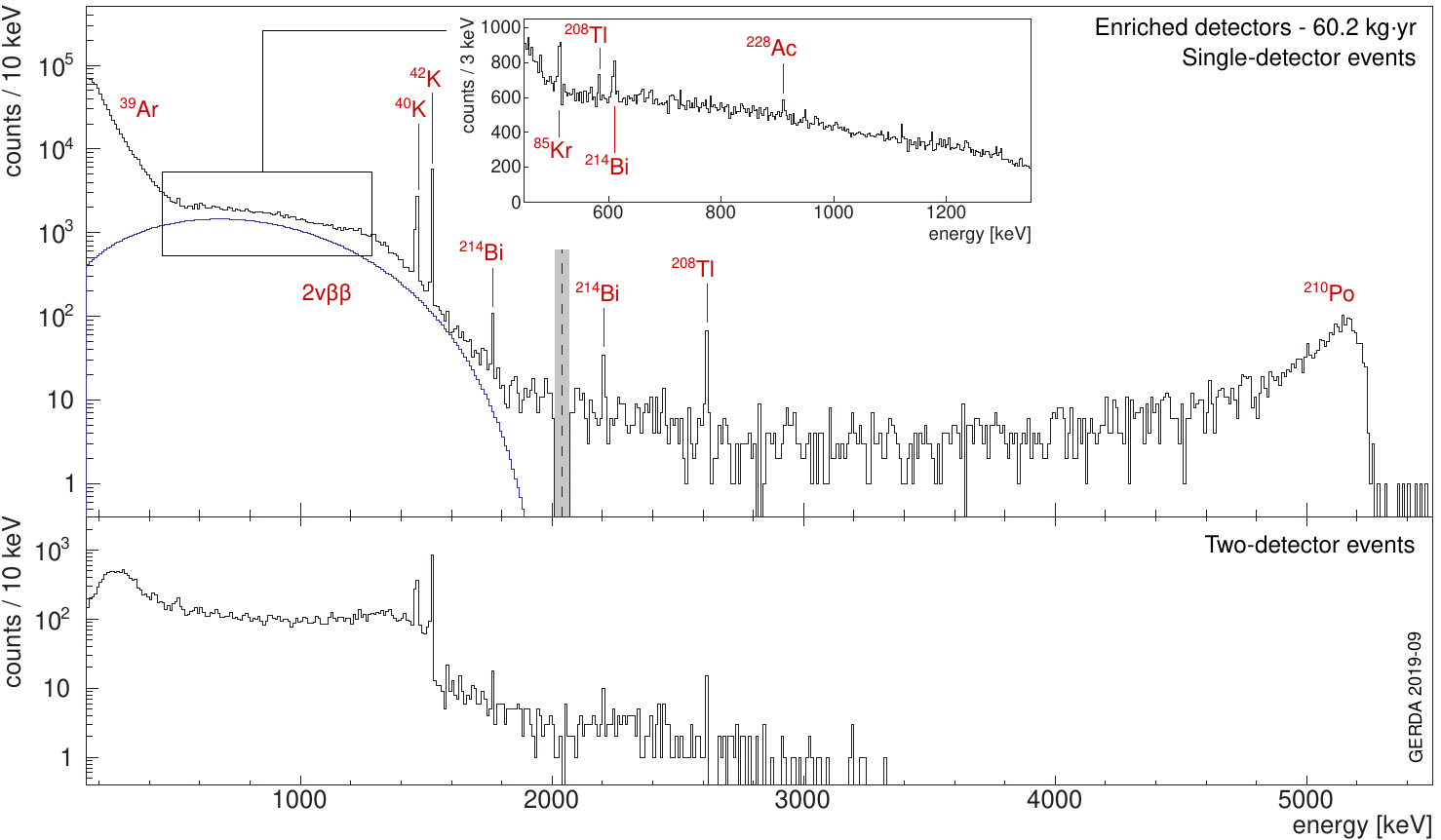}
  \caption{%
    Summed energy spectra of single-detector events (\enrBEGe\ and
    \enrCoax, top panel) and two-detector events (\enrGe, bottom panel)
    collected in \gerdatwo. The prominent features due to detector
    intrinsic \twonu\ events, \kvz, \Ar\ and \Kr\ in the LAr, \kvn, the
    $^{232}$Th and \Uh\ decay chains are highlighted. The window blinded
    for the \onbb\ analysis ($Q_{\beta\beta} \pm 25$~keV) is marked in
    grey.%
  }\label{fig:datadesc}
\end{figure}

The event energy distribution of the three data sets is displayed in
\autoref{fig:datadesc}; the sum spectrum of \enrBEGe\ and \enrCoax\ in
the top panel and \enrGe\ in the bottom panel. For the single-detector
data, in the top panel, the following features are most noticeable: the
$\beta$-decay of \Ar\ dominates the spectrum up to 565~keV while between
600 and 1500~keV the most prominent component is the continuous spectrum
of \nnbb\ decay of \gesix. Two $\gamma$-lines at 1461 and 1525~keV can
be attributed to \kvn\ and \kvz; further visible $\gamma$-lines
belonging to \Kr, \Tl, \Bih\ and \Ac\ are indicated in the figure. The
highest energies displayed are dominated by a peak like structure
emerging at 5.3~MeV with a pronounced low energy tail. This is a typical
spectral feature of $\alpha$-particles and can, here, be attributed to
\Po\ decay on the thin detector \pplus\ surfaces~\cite{Agostini2014}.
Events above the \Po\ peak belong to $\alpha$-decays emerging from the
\Ra\ sub-chain on the detector \pplus\ surfaces. All these components
contribute also to \enrGe\ except for \Ar, \twonu\ and high energy
$\alpha$-components. This is due to the short range of $\alpha$- (tens
of $\upmu$m) and $\beta$-particles (typically smaller than 1.5~cm) in
LAr and germanium with respect to the distance between detectors which
is of the order of several cm.

\begin{figure}[p] \centering
  \subfloat[%
    \Co, \Pa, \Ac\ contaminations and detector intrinsic \twonu\
    decay.\label{fig:pdfs:gmodel:other}%
  ]{\includegraphics[width=0.45\textwidth]{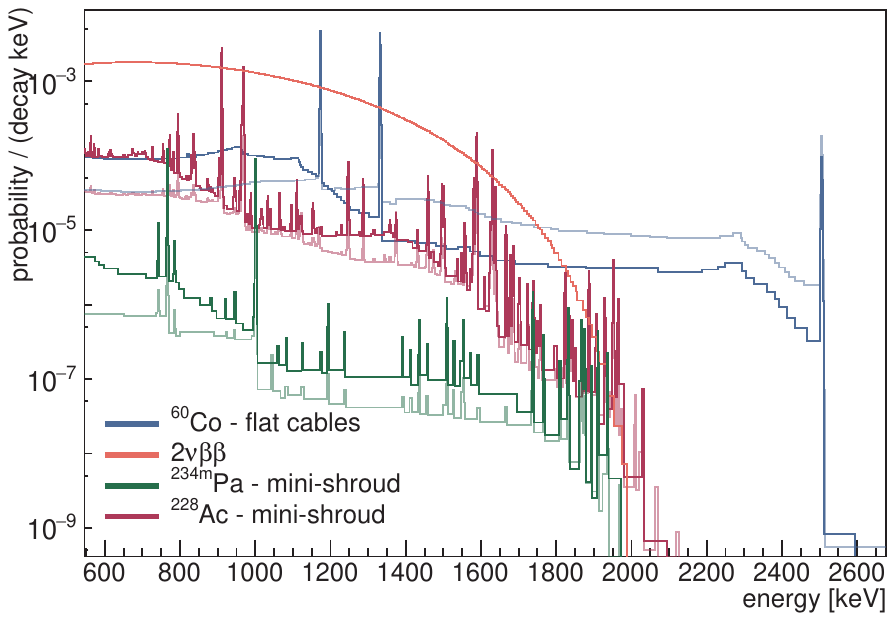}}
  \hspace{10pt}
  \subfloat[%
    \Bil\ and \Tl\ (\Thc\ chain) contaminations far from (fiber shroud)
    and close to (mini-shrouds) the detector
    array.\label{fig:pdfs:gmodel:Th}%
  ]{\includegraphics[width=0.45\textwidth]{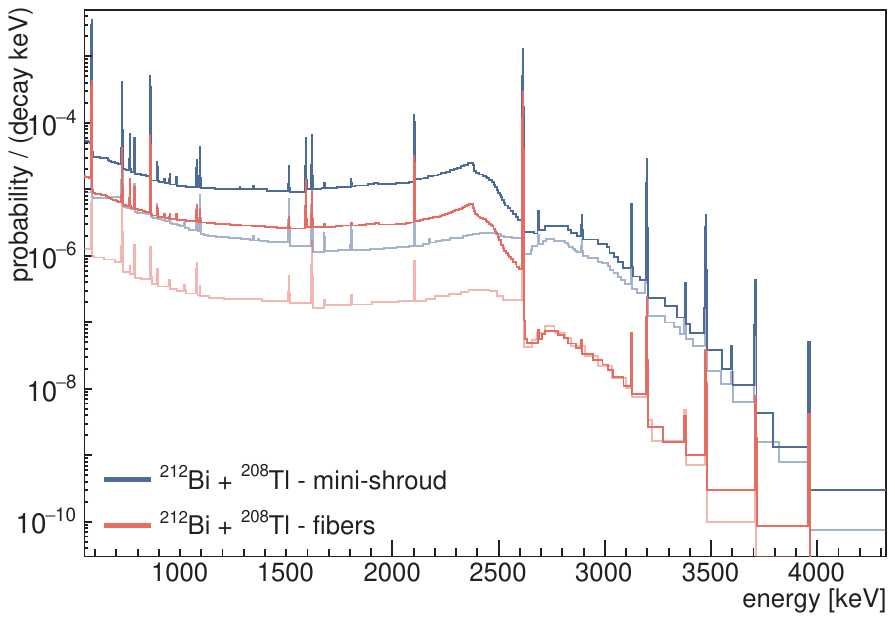}} \\
  \subfloat[%
    \kvn\ contamination close to the detector array (on the
    mini-shrouds), at a higher radial distance (on the fiber shroud) and
    higher vertical distance (on the copper
    shrouds).\label{fig:pdfs:gmodel:K40}%
  ]{\includegraphics[width=0.45\textwidth]{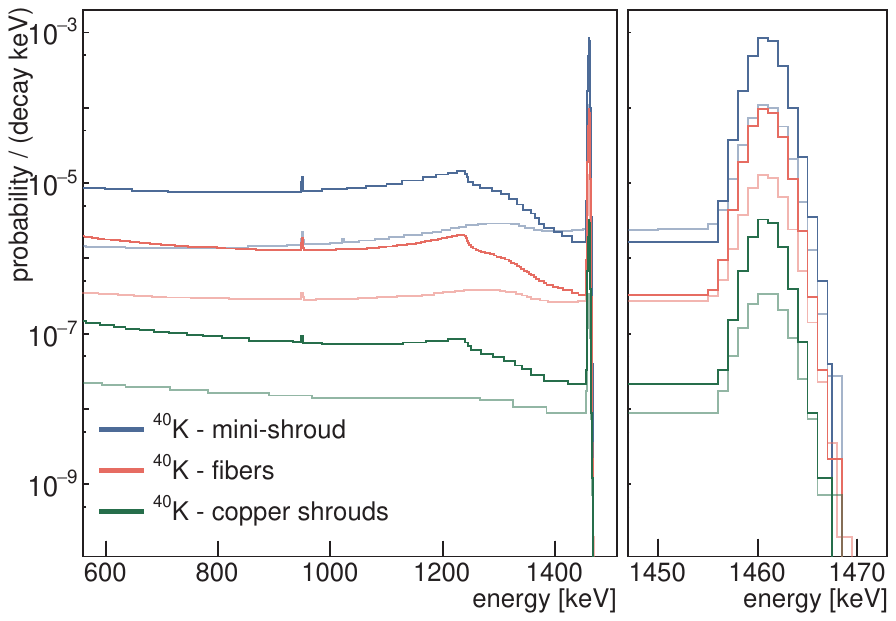}}
  \hspace{10pt}
  \subfloat[%
    \kvz\ contamination in different locations inside the
    LAr.\label{fig:pdfs:gmodel:K42}%
  ]{\includegraphics[width=0.45\textwidth]{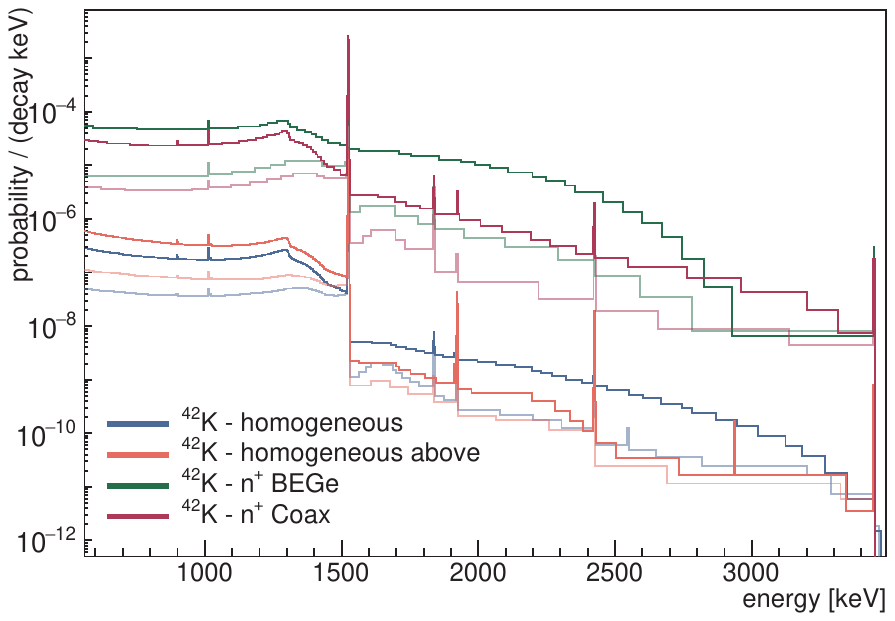}} \\
  \subfloat[%
    \Po\ $\alpha$-decays on \pplus\ contact surface for different
    thicknesses of the inactive contact layer. For 0~nm the nuclear
    recoil energy can be absorbed and some energy can be lost in the
    LAr.
    \label{fig:pdfs:amodel:Po}%
  ]{\includegraphics[width=0.45\textwidth]{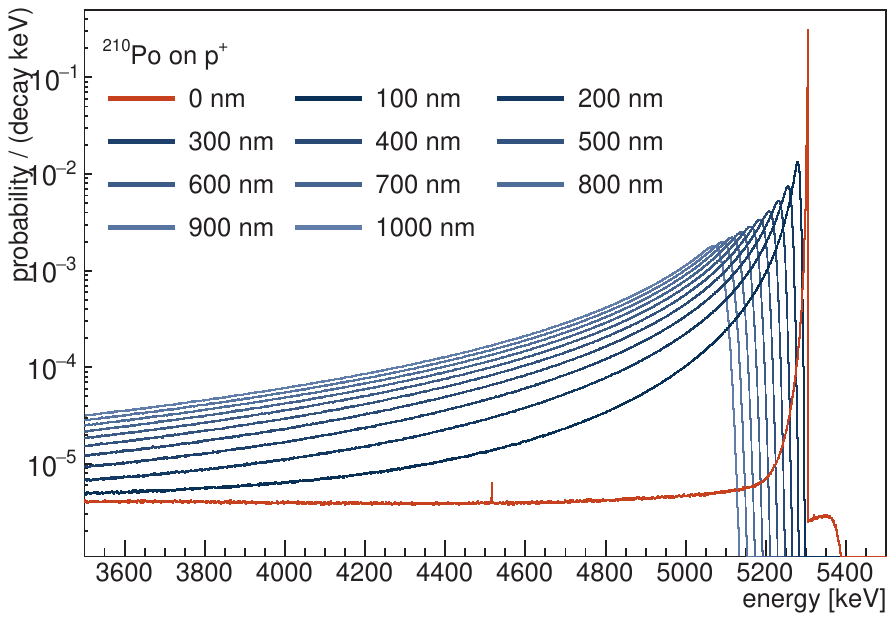}}
  \hspace{10pt}
  \subfloat[%
    \kvz\ contamination in different volumes in the LAr and detector
    intrinsic \twonu\ for comparison. The energy window (ROI) considered
    is \mbox{$(1525\pm4)$~keV} (\kvz\ $\gamma$-line).
    \label{fig:pdfs:kmodel:K42}%
  ]{\includegraphics[width=0.45\textwidth]{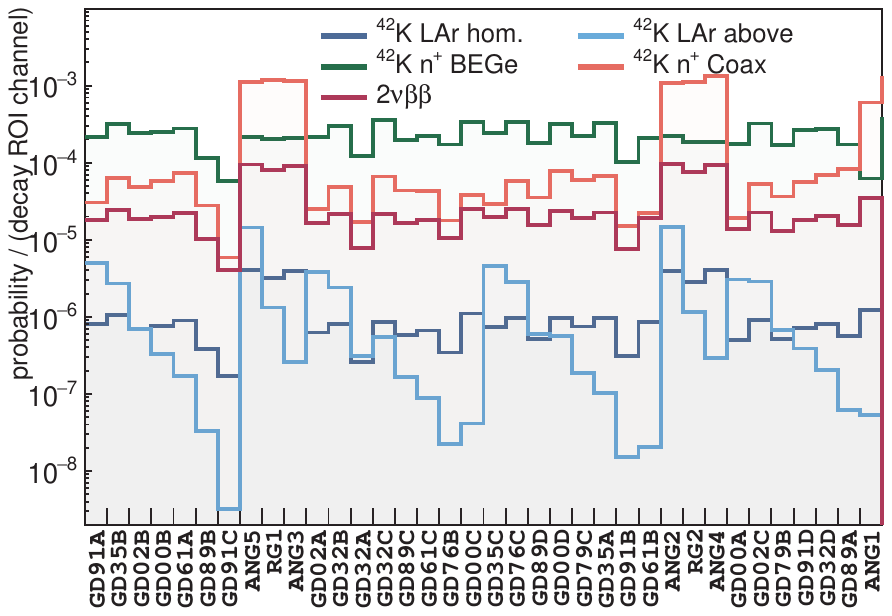}}

  \caption{%
    From (a) to (e): PDFs in the full energy domain. The PDFs for the
    \texttt{M1-enrGe} ($\enrBEGe + \enrCoax$) (in fully opaque colors)
    and the \enrGe\ (in shaded colors) data sets relative to different
    background sources. For visualization purposes a variable binning is
    adopted. (f) PDFs per detector channel for the \kvz $\gamma$-line.
    All PDFs are normalized to the number of simulated primary decays.
  }\label{fig:pdfs:gmodel}
\end{figure}

The structural components of the setup have been screened for their
radio-purity before deployment. Two measurement methods were used
depending on the screened isotope: $\gamma$-ray spectroscopy
(Ge-$\gamma$) with High Purity Germanium (in four underground
laboratories, for details see reference~\cite{Ackermann2013}) and mass
spectrometry with Inductively Coupled Plasma Mass Spectrometers
(ICP-MS)~\cite{Vacri2015}. Especially materials close to the detectors
have been screened for radioactive contaminations originating from the
\Uh\ and \Thc\ decay chains, \kvn\ and \Co. For measured activities and
upper limits see reference~\cite{Agostini2018a} Sec.~5. All possible
background sources taken into consideration in this analysis are
described in detail below. The descriptions are accompanied by a
selection of PDFs in \autoref{fig:pdfs:gmodel} (see also
\autoref{apdx:pdfs}).

\begin{description}
  \item[\Thc\ and \Uh\ decay chains] The only isotopes simulated are
  \Pa, \Pbh\ and \Bih\ from the \Uh\ decay chain and \Ac, \Bil\ and \Tl\
  from the \Thc\ decay chain. The following groups of isotopes are
  assumed to be in secular equilibrium: [\Uh, \Pa] [\Ra, \Pbh, \Bih]
  [$^{228}$Ra, \Ac] and [\Th, \Bil, \Tl]. Their decay products consist
  of $\gamma$- or $\beta$-particles with an energy higher than 520~keV.
  Less energetic particles from the remaining constituents in the chain
  do not enter the energy window which is considered in the presented
  analysis. The $\alpha$-emitters from the decay chains contaminating
  the thin \pplus\ electrodes are described below.

  \item[\Co] A significant fraction of components in the \gerda\ setup
  is made of copper~\cite{Agostini2018a}, which can be produced with
  high radio-purity but is potentially activated by cosmic rays
  producing the long-lived isotope \Co. The latter decays with a
  half-life of 5.2711(8)~yr; from material screening it is also expected
  to be found in some of the detector high-voltage flexible flat cables.

  \item[\kvn] This isotope is found in all screened materials.
  Construction materials were not optimized for ultra-low \kvn\ content
  because the Q-value of its decay is well below \qbb\ and hence does
  not contribute to the background in the ROI. The \kvn\ decay spectrum
  exhibits a $\gamma$-line at 1460.822(6)~keV with an accumulated
  statistics on the order of 100~cts/detector. In
  \autoref{fig:apdx:pdfs:kmodel} the expected counts per detector
  channel for \kvn\ simulated in different locations are shown. Using
  the ratio of events detected in different detectors, information about
  the spatial distribution of \kvn\ can be extracted. We use this
  spatial information to resolve degeneracies of \kvn\ in the energy
  spectra (for details see \autoref{apdx:kmodel}).

  \item[\kvz] A cosmogenically produced isotope in LAr is $^{42}$Ar
  ($T_{1/2} = 32.9(11)$~yr) which decays to \kvz. The distribution of
  \kvz\ inside the LAr is likely to be inhomogeneous due to drift of the
  ionized decay product induced by the electric field (generated by
  high-voltage cables and detectors) and convection.  \kvz\ decays to
  $^{42}$Ca via $\beta$-decay with a half-life of 12.355(7)~h and a
  Q-value of 3525.22(18)~keV, well above \qbb. For the $\beta$-particle
  to be detected the decay needs to happen within a distance of a few
  centimeters\footnote{The path length of \kvz\ $\beta$-particles in LAr
  is less than 1.6~cm, but bremsstrahlung photons from the interaction
  with LAr can travel as far as $\sim$10~cm.} to the detector surface.
  Therefore, we use two distinct PDFs for \kvz\ in LAr generated from
  decays inside and outside the mini-shrouds. As the detectors are in
  direct contact with the LAr, the $\beta$-component of \kvz\
  potentially gives one of the most significant contributions to the
  background in the ROI. A fraction of events around \qbb\ coming from
  \kvz\ is potentially due to $\gamma$-particles with higher energy and
  sub-percent level branching ratio or simultaneous energy deposition of
  multiple $\gamma$-particles.  This $\gamma$-component could become
  important for large quantities of \kvz\ not located directly on the
  detector surfaces with the $\beta$-particle being absorbed in the LAr.
  As for \kvn\ also the $\gamma$-line at 1525~keV of \kvz\ contains
  valuable information about the spatial decay distribution of this
  isotope. In contrast to \kvn\ no additional information, e.g.~from
  radio-purity screening measurements, is available.

  \item[$\alpha$-emitters] The lithium-diffused \nplus\ detector
  surfaces act as a barrier for $\alpha$-particles. The latter can only
  penetrate the very thin boron-implanted \pplus-contact or the contact
  separating groove. $\alpha$-particles have to be emitted directly at
  the surface or from a thin adjacent layer of LAr. Since
  $\alpha$-particles have to cross the $\sim 0.5$~\mum\ thick \pplus\
  dead layer and therefore only part of their initial energy is
  deposited in the active volume, this background component leads to
  peaks with characteristic low-energy tails in the HPGe energy spectra
  (see \autoref{fig:pdfs:amodel:Po}). Some $\alpha$-events, presumably
  originating from the detector groove, are reconstructed with degraded
  energy and lead to an additional, continuous spectral component. We
  find mainly \Po\ but also traces of isotopes from the \Ra\ decay
  chain.

  \item[Detector bulk impurities] Cosmogenically produced long-lived
  isotopes can also be found in
  germanium~\cite{Meierhofer2009,Meierhofer2010,Meierhofer2012}. In
  particular, $^{68}$Ge and \Co\ can occur as detector intrinsic
  impurities with half-lives of 270.93(13)~d and 5.2711(8)~y.  The BEGe
  detectors were kept underground during major parts of the fabrication
  and characterization operations. Periods when these detectors were
  above ground have been tracked in a database~\cite{Agostini2015e}.
  Thus, for the well-monitored BEGe detectors we expect impurities of
  5~nuclei/kg of $^{68}$Ge and 21~nuclei/kg of \Co\ as of September
  2014~\cite{Agostini2015e}. Extrapolating the expected impurities to
  the whole \phasetwo\ data taking period we expect on average
  0.03~cts/day from $^{68}$Ge and 0.1~cts/day due to \Co. From
  background modeling in Phase I~\cite{Agostini2014} the contribution
  for the coaxial detectors formerly used in the Heidelberg-Moscow
  (\hdm)~\cite{Klapdor2001} and \igex~\cite{Aalseth2002} experiments is
  expected to be even smaller due to their long storage underground.
  Simulating the expected detector bulk impurities we find background
  contributions around \qbb\ of less than $10^{-4}$~\ctsper\ in both
  cases. Hence, we conclude that $^{68}$Ge as well as \Co\ can be
  neglected in the following analysis. Potential bulk contaminations
  with \Uh\ and \Thc\ were studied in
  reference~\cite{Collaboration2017}. Only upper limits were found,
  establishing germanium crystals as material of outstanding
  radio-purity. Hence, we only consider the decay of \gesix\ via \nnbb\
  as detector intrinsic background component while all other intrinsic
  impurities are considered to be negligible.

  \item[Other sources] As discussed in reference~\cite{Agostini2014},
  prompt cosmic muon induced background events are efficiently vetoed by
  the identification of \v{C}erenkov light emitted by muons when they
  pass the water tank. The expected BIs, due to the direct muon and
  neutron fluxes at the LNGS underground laboratory, have been estimated
  to be of the order $3\cdot10^{-5}$~\ctsper~\cite{freund2014} and
  $10^{-5}$~\ctsper~\cite{Meierhofer2012} in earlier works,
  respectively. Background contributions coming from delayed decays of
  $^{77}$Ge and $^{77\text{m}}$Ge, also induced by cosmic muons, are
  estimated to be
  $0.21\pm0.01$~nuclei/(kg$\cdot$yr)~\cite{Wiesinger2018} corresponding
  to a BI prior to the active background suppression techniques of about
  $10^{-5}$~\ctsper. Also, the water tank and LAr cryostat
  contaminations are expected to contribute to the \gerda\ BI with less
  than $10^{-4}$~\ctsper~\cite{Ackermann2013,Barabanov2009}. All above
  mentioned contributions are considered negligible in this work. Other
  potential sources of background from interactions of
  \gesix~\cite{Meierhofer2012,vanhoefen2018} and
  $^{206}$Pb~\cite{Mei2008} with neutrons and $^{56}$Co for which no
  evidence was found are not taken into consideration. The
  cosmogenically produced isotope \Ar\ and the anthropogenic isotope
  \Kr~\cite{Winger2005}, which are dissolved in LAr, emit particles
  which are dominantly less energetic than the energy window which is
  considered in the presented analysis.
\end{description}

\afterpage{\clearpage}
\section{Statistical analysis}%
\label{sec:stat-ana}

The multivariate statistical analysis, which is used to model and
disentangle the background in its components, runs on the three binned
data sets \enrBEGe, \enrCoax\ and \enrGe. It is based on the
reconstructed energy with the zero area cusp (ZAC) filter algorithm
which is close to optimal and provides an excellent low-frequency
rejection~\cite{Agostini2015}. The single-detector data sets \enrBEGe\
and \enrCoax\ contain the reconstructed ZAC energy of all \texttt{M1}
events whereas for the two-detector events the sum of the two
reconstructed energies is put in the \enrGe\ data set. Moreover, the
count rate per detector is used for the two potassium $\gamma$-lines.
The spatial event distribution is a collection of the number of events
per detector for \texttt{M1} events and expressed in a matrix of pairs
of detectors for all \texttt{M2} events.

Assuming that the number of events in each bin follows the Poisson
probability distribution $\text{Pois}(n;\nu)$, where $\nu$ is the
expected mean and $n$ is the experimentally measured number of counts,
the likelihood function for a binned data set reads
$\prod_{i=1}^{N_\text{bins}} \text{Pois}(n_i;\nu_i)$. Here $\nu_i =
\sum_{k=1}^{N_\text{com}} \nu_i^{(k)}$ is the expected number of events
in the $i$-th bin, calculated as the sum of the contributions from each
background component $k$; $\nu_i(\lambda_1, \ldots,
\lambda_{N_\text{com}})$ is a
function of the parameters of interests $\lambda_j$ (isotope activities,
\nnbb\ half-life, etc.). The complete likelihood function adopted for
the present analysis combines the three data sets \enrBEGe, \enrCoax\
and \enrGe:
\begin{equation}\label{eq:likelihood}
  \mathcal{L}(\lambda_1, \ldots, \lambda_m \,|\, \text{data}) =
    \prod_{d=1}^{N_\text{dat}}
    \prod_{i=1}^{N_\text{bins}}
    \text{Pois}(n_{d,i};\nu_{d,i})\;.
\end{equation}

The statistical inference is made within a Bayesian framework. Hence, to
obtain posterior probabilities for the free parameters of interest
$\lambda_j$, the likelihood defined in \autoref{eq:likelihood} is
multiplied according to the Bayes theorem by a factor modeling the prior
knowledge of each background component as presented in
\autoref{subsec:priors}. The computation is performed using a Markov
Chain Monte Carlo (MCMC) and is implemented using the BAT software
suite~\cite{Caldwell2009,Beaujean2018}. Posterior probability
distributions of any observable that is not a free parameter of the
likelihood function, like background index estimates, are obtained by
sampling the desired parameter from the MCMC. A \pvalue\ estimate is
provided as a goodness-of-fit measure by adopting the algorithm
suggested in reference~\cite{beaujean2010} for Poisson-distributed data.
It has to be kept in mind that this \pvalue\ estimate, however, is not
as well suited for model comparison as is for instance a Bayes factor;
e.g.~the number of free parameters is not taken into account while a
Bayes factor always penalizes models that add extra complexity without
being required by the data.

\subsection{Analysis window and binning}

The fit range and data bins are chosen such as to exploit as much
information from spectral features as possible brought by data without
introducing undesired bias. The chosen fit range in energy space for the
single-detector data sets (\enrBEGe\ and \enrCoax) starts from just
above the end-point of the \Ar\ $\beta^-$-spectrum at 565~keV and ends
just above the \Po\ peak at 5260~keV, where the event rate drops to
almost zero values. For the two-detector events (\enrGe\ data set) the
fit range starts at 520~keV and extends up to 3500~keV. Possible
additional components outside of this range (e.g. \Ar) do neither add
information to the background decomposition in the ROI around \qbb\ nor
to the analysis of \twonu\ decay. Furthermore, at energies lower than
$\sim$100~keV the shape of the PDFs is dominated by uncertainties on the
detector transition layer model, which describes the charge-carrier
collection at the interface between the \nplus\ contact and the detector
active volume. The exact nature of this transition region is different
for each detector and prone to systematic
uncertainties~\cite{lehnert2016}. 

With an energy resolution which is typically 3--4~keV at \qbb\
(FWHM)~\cite{Agostini2018, Agostini2019} and better at lower energies, a
fixed bin size of 1~keV was chosen for all data sets. The only
exceptions are the two $\gamma$-lines from \kvn\ and \kvz\ each of which
is combined in a single bin from 1455~keV to 1465~keV and from 1520~keV
to 1530~keV, respectively. This is done in order to suppress any
systematic uncertainties of the energy calibration and resolution model
that affect the position and shape of the
$\gamma$-lines~\cite{Agostini2019}.

\subsection{Likelihood factorization}\label{subsec:likelihoodfact}

A feature of the selected data is that the likelihood in
\autoref{eq:likelihood} can be factorized in uncorrelated parts which
can be studied individually and in detail. In the following we shortly
outline the parts of the data which were studied in depth based on the
approach of factorizing the likelihood into uncorrelated parts. Finally,
the results of these analyses are incorporated into a full-range
fit. This procedure is equivalent to a simultaneous analysis of all data
but increases the input knowledge for the fit and breaks down the
computational complexity in smaller steps.

\subsubsection{Potassium tracking analysis}
\label{subsec:stat:klinesshort}

As can be noted from \autoref{fig:pdfs:gmodel:K40} and
\autoref{fig:pdfs:gmodel:K42} the PDFs of \kvn\ and \kvz\ in energy
are prone to degeneracies and hence parameter correlations. Their most
prominent $\gamma$-lines at 1461~and 1525~keV, respectively, contain
information on the spatial distribution while the two-detector events
contain information about the angular distribution of Compton scattered
events. Their combination is beneficial in order to pin down the
potential location of the two potassium isotopes. In total the
\texttt{M1} data contains 4472~cts in $1461\pm4$~keV and 6718~cts in
$1525\pm4$~keV while the \texttt{M2} events contain 554~cts in
$1461\pm6$~keV and 865~cts in $1525\pm6$~keV, respectively. An analysis
of the number of events in the two potassium $\gamma$-lines in each
detector (and detector pair) is used to exploit mainly top-down and
rotational asymmetries in the \kvn\ and \kvz\ distributions. The number
of events in the two energy windows are summarized detector-by-detector;
in the following we refer to this procedure as \textit{projection in
detector space}. The treatment of the likelihood in
\autoref{eq:likelihood} is outlined in detail in \autoref{apdx:kmodel}.
The number of events in all other $\gamma$-lines is too low in order to
adopt a useful detector-wise analysis. The spatial analysis of \kvn\ and
\kvz\ is incorporated in the full-range fit by directly employing the
posterior parameter distributions as prior information.\footnote{By
adopting this approach, a part of the data in the potassium
$\gamma$-lines region is analyzed twice: first in the potassium tracking
analysis and then in the full-range fit. Nevertheless, considering that
the two analyses exploit different data features (i.e.~count rate per
detector and total count rate per energy) and the overlap between the
two data set is minimal, the overall effect is negligible.}

\subsubsection{\texorpdfstring{$\alpha$-events background analysis}{a-events background analysis}}
\label{subsec:stat:alphashort}

The single-detector energy spectra above 3.5~MeV (the Q-value of \kvz\
$\beta$-decay) are strongly dominated by $\alpha$-events. They are not
present in two-detector data due to the short range of
$\alpha$-particles in LAr and germanium. Also, this component is not
correlated to other backgrounds considered here because it peaks at
energies well above the highest $\gamma$-emission energies and
$\beta$-decay Q-values. A careful study was carried out considering
various \pplus\ contact thickness and event rates to reproduce the \Po\
peak. In order to reproduce $\alpha$-events with degraded energy an
empirical model is fit to the data. A linear function with free slope
and offset and a cut-off below the maximum of the \Po\ peak fits the
data well. The agreement of the $\alpha$-background model with the data
is demonstrated in \autoref{apdx:amodel} and \autoref{fig:apdx:alphafit}
therein. Information from the detailed analysis of the high-energy
$\alpha$-region is incorporated in the full-range fit using a combined
PDF that summarizes the \Po\ peak plus the \Ra\ decay chain and a linear
floating component for degraded $\alpha$-events.

\subsection{Prior distributions}

The following criteria are adopted to convert the prior information
described in \autoref{subsec:priors} into prior probability
distributions on the parameters of interest\footnote{In Bayesian
analysis the prior probability distribution describes all knowledge
about an unobserved quantity of ultimate interest before taking the data
into account.}: if a measured value with uncertainty is available for a
background contamination then a Gaussian distribution with a
corresponding centroid and a $1\sigma$ width is adopted. In presence of
a 90\% C.L.~upper limit, instead, an exponential prior distribution is
constructed with 90\% of its area covering parameter values from 0 up to
the given 90\% C.L.~upper limit. A uniform prior distribution is
assigned to components for which no measured value or upper limit is
available. Ranges for uniform priors are initially taken very wide, in
order to span a large portion of the allowed parameter space, then
optimized to contain at least 99\% of the posterior distribution. As
mentioned before, in addition to the information from screening
measurements, prior distributions for \kvn\ and \kvz\ are constructed
considering the posterior inference from their spatial
distribution.\footnote{The Bayesian posterior distribution is the
conditional probability distribution of the unobserved quantities of
ultimate interest, given the observed data.} Moreover, as \Bih\ is part
of the \Ra\ decay chain, we constrain a \Bih\ component on the \pplus\
contact by a Gaussian prior extracted from the obtained \Ra\ activity
based on the energy estimator in the high-energy $\alpha$ region.


\section{Results}%
\label{sec:results}

As described in \autoref{subsec:likelihoodfact} the $\alpha$-event
background and potassium $\gamma$-lines are studied individually and the
results are incorporated in the full-range fit as prior distributions.
The latter combines a simultaneous fit of the \texttt{M1} and the
\texttt{M2} data sets. For the final combination of parameters, outlined
in this section, components with a posterior distribution peaked at zero
were eliminated from the fit. The stability of the results with respect
to the bin size and prior distributions was verified. Changing the prior
distribution for fit parameters for which no screening measurement is
available from a flat to an exponential one does not significantly
impact the final posterior distributions. The compatibility of the final
model, which includes 34 free fit parameters, with data is supported by
a \pvalue\ of $\sim$~0.3.

The estimated activities of individual components and other parameters
of interest are listed in \autoref{tab:parameter_estimates}. In
particular, for each component we report the global and the marginalized
mode of the posterior parameter distribution, along with its smallest
68\% C.I. The global mode corresponds to the global best fit value
while the marginalized mode is the most probable parameter value when
integrating over all other parameters. The original type of prior
distribution is marked with \m{[f]} for flat, \m{[g]} for Gaussian and
\m{[e]} for exponential; the latter two are used if screening
measurements are available. Subsequently, for all \kvn\ and \kvz\
components, the prior distribution is imported from the potassium
tracking analysis and for \Pbh\ and \Bih\ on the p$^+$ contact from the
reconstructed \Ra\ content from the $\alpha$-events background analysis.

\begin{figure}[tb]
  \centering
  \includegraphics{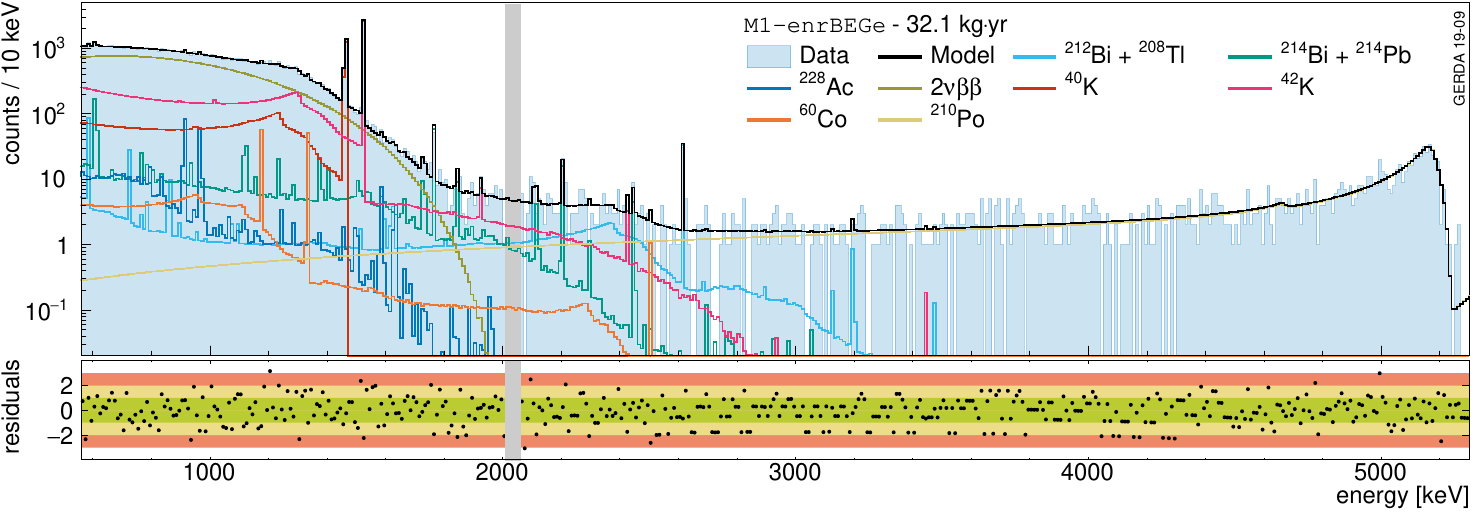}
  \includegraphics{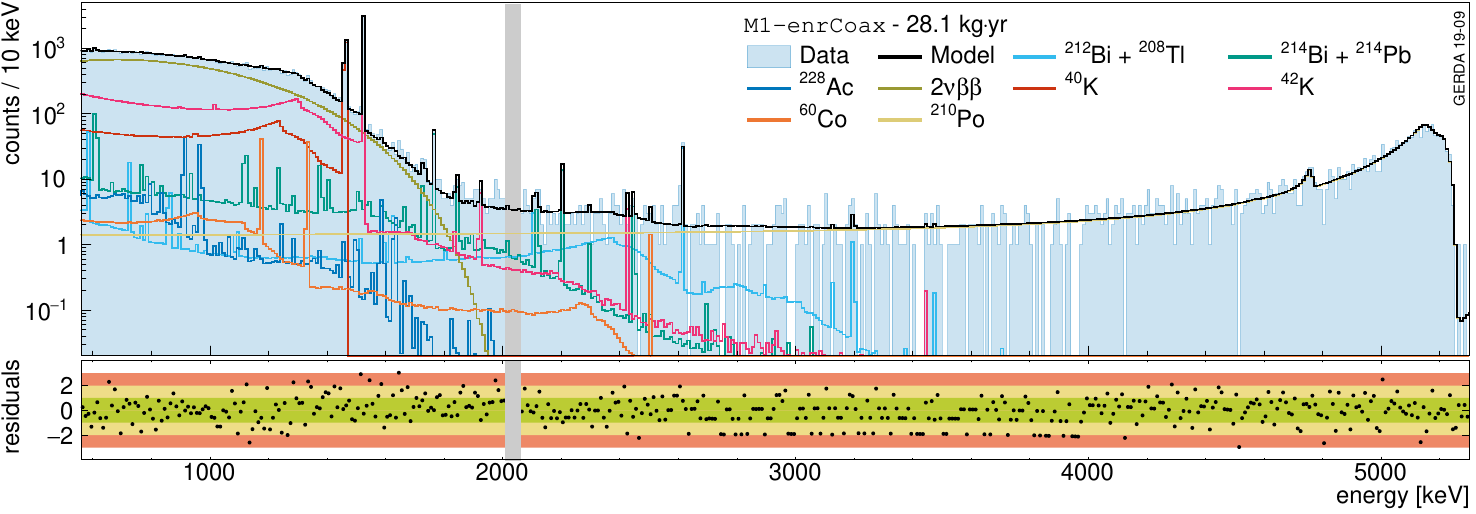}
  \includegraphics{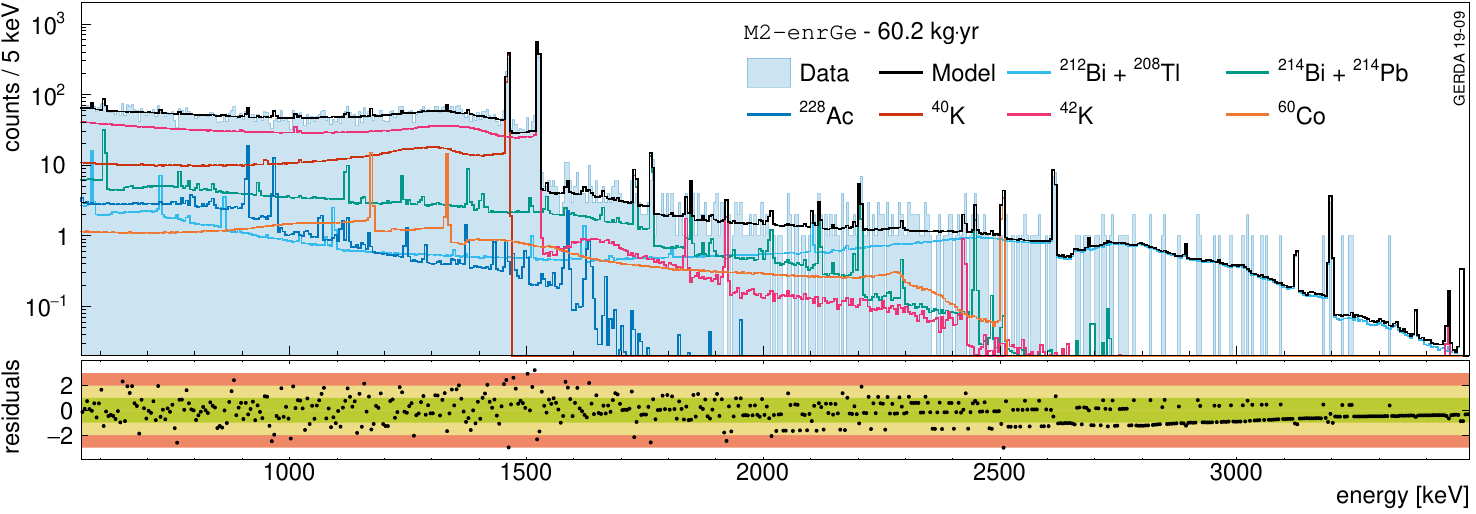}
  \caption{%
    Background decomposition of the event energy distributions of the
    (from top to bottom) \enrBEGe, \enrCoax\ and \enrGe\ data sets.
    Components referring to the same background source in different
    locations are summed together for visualization convenience. The
    blinded region $Q_{\beta\beta}\pm25$~keV is highlighted in gray. In
    the three lower panels displaying the normalized residual
    distributions the central 1$\sigma$-, 2$\sigma$- and 3$\sigma$-bands
    are marked in green, yellow and red, respectively. Note that for
    bins with low expected statistics due to the discrete nature of the
    measured spectrum not all colored bands are
    meaningful~\cite{Aggarwal2012}.%
  }\label{fig:gmodelspc}
\end{figure}

The spectral decomposition of all data sets is shown in
\autoref{fig:gmodelspc}. For each data set the residual distribution as
a multiple of the expected 1$\sigma$ fluctuation in each bin is
displayed. We find for the \enrBEGe\ data set 66.4\%, 94.5\% and 99.6\%
of points in the 1$\sigma$-, 2$\sigma$- and 3$\sigma$-bands, for the
\enrCoax\ data set 66.0\%, 94.7\% and 99.8\% and for the \enrGe\ data
set 70.0\%, 96.1\% and 99.7\%, respectively. Thus, in all three cases
the residuals are normally distributed. No outliers with residuals
larger than $3\sigma$ are found in a $\pm50$~keV window around \qbb\ and
the bins exceeding $3\sigma$ do not correspond to any noted
$\gamma$-line.

\afterpage{%
  \clearpage%
  \thispagestyle{empty}
  \begin{landscape}
    \small
    \centering
    \newcolumntype{H}{>{\setbox0=\hbox\bgroup}c<{\egroup}@{}}
\newcolumntype{x}[1]{>{\centering\arraybackslash}p{#1}}
\newcommand{\mr}[2]{\multirow{#1}{*}{#2}}
\newcommand{\mrr}[2]{\multirow{#1}[1]{*}{#2}}
\newcommand{\mrc}[2]{\multirowcell{#1}{#2}}
\begin{tabular}{%
  r
  l
  c
  S[
    table-format=4.2,
    table-number-alignment=right,
    table-text-alignment=right,
    table-parse-only
  ]
  r@{ }l
  S[
    table-number-alignment=left,
    table-text-alignment=left,
    table-parse-only
  ]
  S[
    table-format=5.0,
    table-number-alignment=center,
    table-text-alignment=center,
    table-auto-round=true
  ]@{ | }
  l
  S[
    table-format=4.0,
    table-number-alignment=center,
    table-text-alignment=right,
    table-auto-round
    ]@{ | }
  l
  S[
    table-format=4.0,
    table-number-alignment=center,
    table-text-alignment=center,
    table-auto-round
  ]
}
  \toprule
  \mrr{3}{source}               & \mrr{3}{\m{[prior]} location}         & \mrr{3}{units}  & {\mrc{3}{global\\mode}} & \multicolumn{2}{c}{\mrc{3}{marg.~mode\\with 68\% CI}} & {\mrr{3}{screening}} & \multicolumn{5}{c}{model content in fit range | BI at \qbb} \\
                                &                                       &                 &                         &                                                     & &                     & \multicolumn{5}{c}{units: cts | $10^{-3}$\ctsper}                \\
  \cmidrule(lr){8-12}
                                &                                       &                 &                         &                                                     & &                     & \multicolumn{2}{c}{\enrBEGe} & \multicolumn{2}{c}{\enrCoax} & \enrGe  \\
  \midrule
  \mr{2}{\nnbb}                    & \m{[f]} germanium                     & $10^{21}$yr       & 2.025 & 2.030 & $[2.016, 2.044]$ & {--}        & 45272.0 & \mrc{2}{0}                   & 37866.7 & \mrc{2}{0}                   & {--}    \\
                                   & \m{[f]} $\delta^{2\nu}$ (Coax)        & cts               & 2890  & 3200  & $[2600, 3600]$   & {--}        & {--}    &                              & 1962.2  &                              & {--}    \\
  \midrule
  \mr{3}{\Bil\ + \Tl}              & \m{[e]} flat cables                   & \mr{3}{$\upmu$Bq} & 384   & 380   & $[355, 408]$     & < 410       & 424.4   & \mrc{3}{3.52\\$[3.30,3.76]$} & 274.4   & \mrc{3}{2.21\\$[2.03,2.34]$} & 449.1   \\
                                   & \m{[g]} copper shrouds $^{\dagger}$   &                   & 194   & 197   & $[175, 213]$     & 194 \pm 19  & 2.9     &                              & 3.1     &                              & 1.4     \\
                                   & \m{[g]} mini-shrouds                  &                   & 18.7  & 17.7  & $[13.8, 23.8]$   & 18  \pm 5   & 20.6    &                              & 21.0    &                              & 24.3    \\
  \midrule
  \mr{6}{\Pbh\ + \Bih}             & \m{[f]} \pplus\ (BEGe)                & \mr{6}{$\upmu$Bq} & 0.36  & 0.35  & $[0.27, 0.53]$   & {--}        & 5.8     & \mrc{6}{2.63\\$[2.50,2.78]$} & 0.0     & \mrc{6}{3.16\\$[2.83,3.50]$} & 2.9     \\
                                   & \m{[f]} \pplus\ (Coax)                &                   & 1.053 & 1.07 & $[0.91, 1.30]$ & {--}        & 0.0     &                              & 26.0    &                              & 5.4     \\
                                   & \m{[g]} flat cables                   &                   & 560   & 552   & $[523, 594]$     & 660 \pm 210 & 1193.5  &                              & 750.0   &                              & 923.2   \\
                                   & \m{[g]} copper shrouds $^{\dagger}$   &                   & 533   & 535   & $[480, 585]$     & 532 \pm 53  & 9.1     &                              & 9.9     &                              & 3.9     \\
                                   & \m{[g]} mini-shrouds                  &                   & 45    & 47    & $[33, 59]$       & 43  \pm 13  & 97.5    &                              & 95.8    &                              & 82.6    \\
                                   & \m{[g]} SiPM-ring                     &                   & 353   & 345   & $[256, 450]$     & 351 \pm 97  & 6.4     &                              & 4.9     &                              & 2.9     \\
  \midrule
  \mr{9}{\kvn}                     & \m{[g]} flat cables                   & \mr{9}{mBq} & 2.95  & 2.9   & $[2.1, 4.1]$     & 6  \pm 2    & 861.1   & \mrc{9}{0}                   & 530.0   & \mrc{9}{0}                   & 338.8   \\
                                   & \m{[g]} front-end electronics         &                   & 16.6  & 16.0  & $[11.5, 20.3]$   & 13 \pm 4    & 103.8   &                              & 79.3    &                              & 45.7    \\
                                   & \m{[g]} copper shrouds $^{\dagger}$   &                   & 18.4  & 18.2  & $[16.6, 20.2]$   & 18 \pm 2    & 41.5    &                              & 44.9    &                              & 17.3    \\
                                   & \m{[g]} fiber shroud                  &                   & 2.73  & 2.83  & $[2.29, 3.39]$   & 2.9 \pm 0.6 & 124.2   &                              & 116.2   &                              & 55.1    \\
                                   & \m{[g]} detector holders              &                   & 1.64  & 1.75  & $[1.29, 2.07]$   & 2.8 \pm 0.6 & 885.8   &                              & 468.1   &                              & 334.2   \\
                                   & \m{[g]} mini-shrouds                  &                   & 1.70  & 1.69  & $[1.60, 1.80]$   & 1.7 \pm 0.6 & 517.6   &                              & 474.9   &                              & 215.7   \\
                                   & \m{[g]} SiPM ring                     &                   & 1.95  & 3.0   & $[1.1, 4.4]$     & 2 \pm 2     & 4.7     &                              & 3.6     &                              & 2.1     \\
                                   & \m{[f]} far from the array            &                   & {--}  & {--}  &                  & {--}        & 783.5   &                              & 847.0   &                              & 326.6   \\
                                   & \m{[f]} close to the array            &                   & {--}  & {--}  &                  & {--}        & 3469.1  &                              & 3182.4  &                              & 1445.6  \\
  \midrule
  \mr{4}{\kvz}                     & \m{[f]} \nplus\ (BEGe)                & \mr{2}{$\upmu$Bq} & 261.5 & 295.0 & $[224.3, 324.7]$ & {--}        & 920.2   & \mrc{4}{5.69\\$[4.58,6.29]$} & {--}    & \mrc{4}{1.29\\$[1.15,1.40]$} & 162.4   \\
                                   & \m{[f]} \nplus\ (Coax)                &                   & 490.0 & 415.0 & $[309.6, 506.0]$ & {--}        & {--}    &                              & 805.6   &                              & 162.4   \\
                                   & \m{[f]} LAr {--} above array          & \mr{2}{Bq}        & 0.451 & 0.453 & $[0.437, 0.468]$ & {--}        & 5859.0  &                              & 4421.1  &                              & 2535.0  \\
                                   & \m{[f]} LAr {--} outside mini-shrouds &                   & 2.026 & 2.027 & $[1.985, 2.068]$ & {--}        & 10225.0 &                              & 9691.0  &                              & 4543.8  \\
  \midrule
  \mr{3}{\Ac}                      & \m{[g]} copper shrouds $^{\dagger}$   & \mr{4}{$\upmu$Bq} & 62.0  & 62.5  & $[56.0, 67.9]$   & 62 \pm 6    & 0.5     & \mrc{4}{0.36\\$[0.31,0.40]$} & 0.6    & \mrc{4}{0.33\\$[0.28,0.37]$} & 0.2     \\
                                   & \m{[e]} detector holders              &                   & 183   & 182   & $[158, 208]$     & < 250       & 541.1   &                              & 280.5   &                              & 347.0   \\
                                   & \m{[g]} mini-shrouds                  &                   & 18.0  & 17.8  & $[12.9, 22.8]$   & 18  \pm  5  & 28.0    &                              & 27.4    &                              & 19.5    \\
  \Co                              & \m{[e]} flat cables                   &                   & 113   & 114   & $[98, 130]$      & < 250       & 381.7   &                              & 240.1   &                              & 332.8   \\
  \midrule
  \mr{4}{$\alpha$-decays}          & \m{[f]} \Po\ + \Ra\ chain (BEGe)      & \mr{4}{cts}       & 1173  & 1183  & $[1127, 1253]$   & {--}        & 561.0   & \mrc{4}{3.31\\$[3.12,3.78]$} & {--}    & \mrc{4}{4.76\\$[4.40,5.08]$} & {--}    \\
                                   & \m{[f]} \Po\ + \Ra\ chain (Coax)      &                   & 3320  & 3300  & $[3200, 3400]$   & {--}        & {--}    &                              & 1584.6  &                              & {--}    \\
                                   & \m{[f]} energy-degraded (BEGe)        &                   & 595   & 628   & $[583, 680]$     & {--}        & 587.4   &                              & {--}    &                              & {--}    \\
                                   & \m{[f]} energy-degraded (Coax)        &                   & 700   & 698   & $[641, 747]$     & {--}        & {--}    &                              & 623.3   &                              & {--}    \\
  \bottomrule%
\end{tabular}%

%
    \captionof{table}{%
      Summary of the analysis parameter estimates. Global mode and
      marginalized mode, along with its smallest 68\% C.I., are reported
      as representatives of the posterior parameter distribution. The
      number of reconstructed counts in the fit range and the BI at \qbb\
      prior active background suppression are listed for each component
      and each analysis data set. The original type of prior distribution
      is marked with \m{[f]} for flat, \m{[g]} for Gaussian and \m{[e]}
      for exponential. ($\,^{\dagger}$ \tetratex-coated) %
    }\label{tab:parameter_estimates}
  \end{landscape}
  \clearpage
}

The \kvz\ distribution is optimized to best fit the data. In order to
disentangle the \kvz\ $\gamma$- and $\beta$-components, the volume
inside and outside of the mini-shrouds is separated in the PDF
construction. Inside the mini-shrouds a homogeneous distribution is
compatible with the data as well as \kvz\ attached to the detectors
contact surfaces. In the fit model given here, a possible scenario is
chosen where all \kvz\ is located on the \nplus\ surfaces. However, we
note that \kvz\ on the \pplus\ appears to partly substitute the
energy-degraded $\alpha$-component in the \enrCoax\ data set if
introduced in the fit and predicts a higher total BI around \qbb. The
extracted \kvz\ activity on the $^{\text{enr}}$Coax \pplus\ contact in
this case is $22\pm4~\upmu$Bq corresponding to a contribution to the BI
around \qbb\ of $(7\pm1)\cdot10^{-3}~$\ctsper. For the \enrBEGe\ data
set the posterior distribution of a possible \kvz\ component on the
\pplus\ contact is compatible with zero. Outside the mini-shrouds an
inhomogeneous distribution of the \kvz\ decays explains the observations
better. Detectors which are located at higher positions in the strings
show an excess of events in the \kvz\ 1525~keV $\gamma$-line which is
compatible with a surplus of \kvz\ located right above the detector
array (see \autoref{apdx:kmodel}). The full-range fit model contains a
homogeneous \kvz\ distribution \textit{outside the mini-shrouds} which
is reconstructed with a specific activity of $186\pm39~\upmu$Bq/kg plus
an additional distribution in the vicinity of the cables \textit{above
the array}.

A large fraction of the contamination with \kvn\ in the setup can not be
accounted for by the screened hardware listed in
\autoref{tab:parameter_estimates}. We thus add a close ($\sim1$~cm) and
a far ($\sim50$~cm) \kvn\ component with respect to the detector array
which are in fact replica of the PDFs for the mini-shrouds and the
\tetratex-coated copper shrouds. These additional components absorb the
excess indicated by the fit, the largest part of the reconstructed
events in the spectra is attributed to impurities close to the array.

The \kvn\ and \kvz\ distributions can be further split into smaller
volumes and studied as an extension of the potassium tracking analysis
(as described in \autoref{subsec:stat:klinesshort}) \textit{projected in
detector space}. The additional \kvn\ component close to the array and
the \kvz\ component above the array are split into 7 sub-components on a
string-by-string basis. The potassium concentration is in general found
to be asymmetric among the detector strings. In particular, a more
prominent \kvz\ concentration is found above the central string. This is
consistent with the electrostatic drift of \kvz\ ions induced by the
electric field in the LAr which is generated by the unshielded
high-voltage flat cables biased with about 4~kV. The \kvn\ and \kvz\
spatial analysis fitting the potassium $\gamma$-lines \textit{projected
in detector space} is presented in full detail in \autoref{apdx:kmodel}.

The $\alpha$ distribution is adjusted to best fit the data. The \Po\
peak at 5.2~MeV is found to be best described by a mixture of PDFs
obtained assuming different \pplus\ contact thicknesses confirming
results of the \phaseone\ background analysis~\cite{Agostini2014}. The
empirical linear model which is used to describe $\alpha$-events with
degraded energy (see \autoref{subsec:stat:alphashort}), extends down to
\qbb\ and below. For the \enrBEGe\ data set $\alpha$-events are
efficiently isolated using pulse shape discrimination (PSD) techniques.
The compatibility of the degraded-energy $\alpha$-component with
$\alpha$-events identified by PSD was checked and is found consistent.
All details about the $\alpha$-events analysis can be found in
\autoref{apdx:amodel}. 

Smaller contributions to the background model in the full energy range
are attributed to \Pbh\ and \Bih\ from the \Uh\ decay chain, \Ac, \Bil\
and \Tl\ from the \Thc\ decay chains and \Co. With a total contribution
in the fit range of $10^{-3}$~cts/keV for both the \enrBEGe\ and
\enrCoax\ data set \Pa\ gives negligible contribution to the spectra and
is therefore dropped from the full-range fit model. The central
values preferred in the full-range fit are driven by screening
measurements and the spectral contributions are all fully accounted for
by the listed hardware components. The only exception is \Pbh\ and \Bih\
where a minor contribution is added on the p$^+$ contact expected from
the observation of $\alpha$-events belonging to the \Ra\ decay chain.

Most counts in the fit range are attributed to the \nnbb\ decay of
\gesix; in fact its continuous distribution dominates the spectrum up to
almost 1.9~MeV. Here, we base the \twonu\ half-life estimate on the
\enrBEGe\ data set only. An additional parameter, $\delta^{2\nu}$,
parametrizes the observed discrepancy to the value solely derived from
the \enrCoax\ data set. The value of $\delta^{2\nu}$ extracted from the
fit amounts to a surplus of 5\% of \twonu\ counts observed in \enrCoax.
It mainly quantifies the systematic biases between the active volume
determination methods of the two detector types. The $^\text{enr}$BEGe
detectors active volume measurements are affected by a smaller
systematic uncertainty than the $^\text{enr}$Coax
detectors~\cite{Agostini2014,GERDAcollaboration2019}. Hence, the
extracted \twonu\ half-life, based on the \enrBEGe\ data set and given
here only with statistical uncertainties, amounts to $T_{1/2}^{2\nu} =
(2.03 \pm 0.02) \cdot 10^{21}$~yr. A detailed discussion follows in
\autoref{sec:discussion}.

\begin{figure}[tb]
  \centering
  \includegraphics[width=0.4\linewidth]{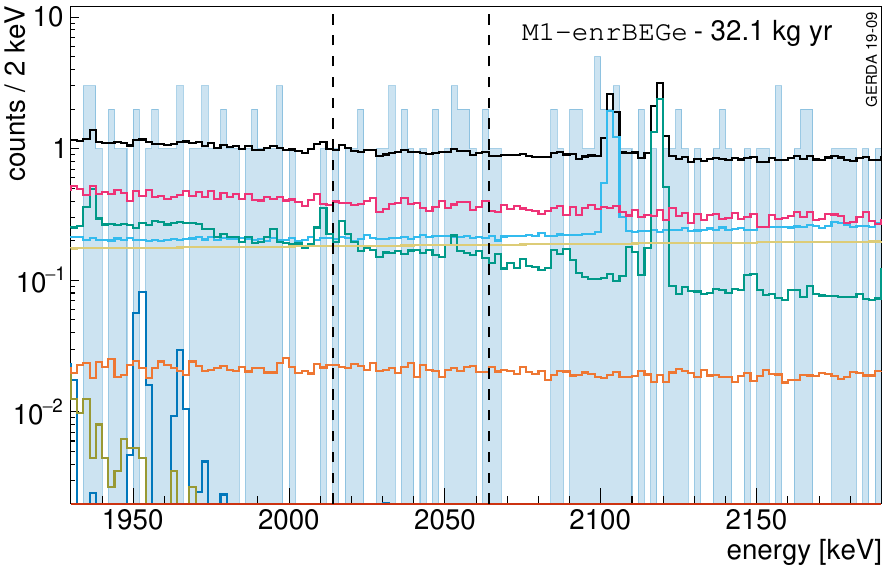}
  \hspace{12pt}
  \includegraphics[width=0.4\linewidth]{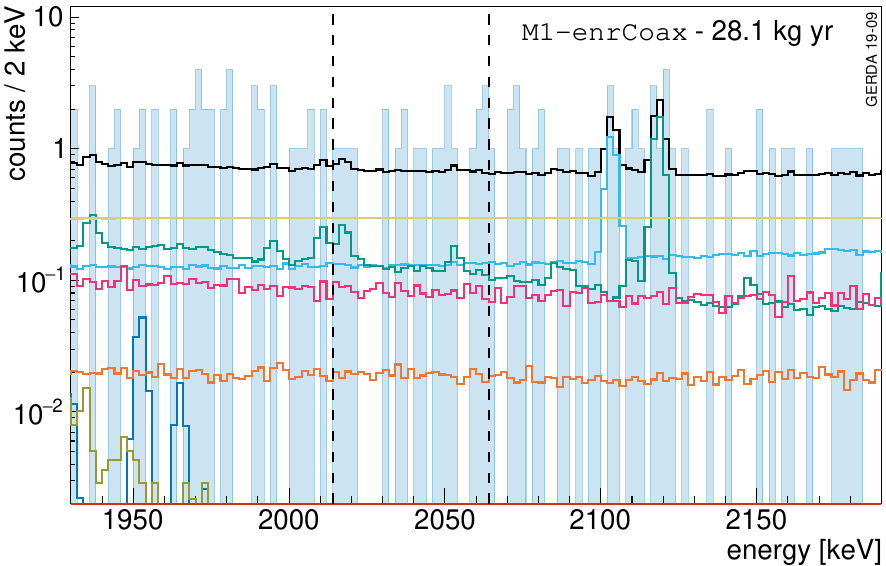}
  \caption{%
    Background decomposition for the \enrBEGe\ (left) and the \enrCoax\
    (right) data sets in the background window between 1930~keV and
    2190~keV after data unblinding. The previously blinded window
    ($Q_{\beta\beta} \pm 25$~keV) is indicated by two dashed lines.
    The background distribution before active background suppression in the
    \onbb\ analysis window can be well approximated with a constant
    function. For color code see \autoref{fig:gmodelspc}.%
  }\label{fig:roishape}
\end{figure}

The background model describes the individual contributions to the total
background index (BI) around \qbb\ prior active background suppression
(see \autoref{fig:roishape}). The BI is defined as the number of counts
over exposure and energy in the energy window from 1930~keV to 2190~keV
excluding the region around \qbb\ ($Q_{\beta\beta} \pm 5$~keV) and the
intervals $2104 \pm 5$~keV and $2119 \pm 5$~keV, which correspond to
known $\gamma$-lines from \Tl\ and \Bih. The values for each background
contribution are given in \autoref{tab:parameter_estimates}. The
dominating background contribution around \qbb\ in the \enrBEGe\ data
set come from \kvz. Isotopes from the \Thc\ decay chain,
$\alpha$-particles mainly with degraded energy and isotopes from the
\Uh\ decay chain contribute about equally. The estimated total BIs
extracted from the marginalized posterior distributions are
$16.04^{+0.78}_{-0.85}\,\text{(stat)} \cdot 10^{-3}$~\ctsper\ for the
\enrBEGe\ data set and $14.68^{+0.47}_{-0.52}\,\text{(stat)} \cdot
10^{-3}$~\ctsper\ for the \enrCoax\ data set.

\section{Discussion}%
\label{sec:discussion}

In general, impurities close to the detector array contribute most to
the background, far components give minor contributions. The posterior
distribution and the screening measurements are in very good agreement
and the spectral content of each source of background can be accounted
for by the screened hardware components. Only in the case of \kvn\ a
large part of the observed activity cannot be explained by the
screened hardware and is fit with the additionally introduced components
far and close to the detector array. The \kvz\ and $\alpha$-event
distributions cannot be constrained by screening measurements and are
adjusted to best fit the data. 

The presented background model is not unambiguous in all components. As
shown in \autoref{fig:pdfs:gmodel} several PDFs of the same source of
background located in different structural components are very similar
and thus prone to correlation. Most of them have been resolved by
introducing prior distributions based on the screening measurements.
However, a few anti-correlations persist which are listed in
\autoref{fig:corr}.

\begin{table}[tb]
  \centering
  \caption{%
    Correlations between fit components
    relative to the same background contamination in different
    locations.
  }\label{fig:corr}
  \begin{tabular}{cccc}
    \toprule
    contamination         & location 1                  & location 2         & correlation \\
    \midrule
    \Bih\ + \Pbh\         & mini-shrouds                & flat cables        & -0.43 \\
    \multirow{2}{*}{\kvn} & flat cables                 & detector holders   & -0.45 \\
                          & flat cables                 & close to the array & -0.63 \\
    \multirow{2}{*}{\kvz} & LAr -- outside mini-shrouds & \nplus\ contact    & -0.42 \\
                          & LAr -- outside mini-shrouds & LAr -- above array & -0.56 \\
    \bottomrule
  \end{tabular}
\end{table}

For what concerns \kvz\ in the LAr volume outside the mini-shrouds and
thus more distant from the detector array, the adopted distribution is
purely empirical. Our prior knowledge is limited by the fact that the
\kvz\ ions undergo drift due to the electrical fields surrounding the
detectors and high-voltage cables. Also, due to thermal gradients they
can be displaced by convection. Hence, their distribution inside the
\gerda\ LAr is prone to systematic uncertainties. The presence of
unshielded high-voltage cables above the detector array can explain the
excess of \kvz\ found in this region. From the perspective of the
full-range fit a more sophisticated modeling does not significantly
modify the \kvz\ PDFs and hence the fit results. A potentially
asymmetric \kvz\ distribution is, thus, not further followed in the main
analysis. Nevertheless, some considerations can be found
in~\autoref{apdx:kmodel}. An explanation for \kvz\ on the \pplus\
contact being rejected for the \enrBEGe\ data set but potentially
present in the \enrCoax\ data can be the specific bore-hole geometry of
the semi-coaxial detectors. \kvz\ produced inside the hole can not
easily escape and is trapped close to the \pplus\ contact. 

For each source of background the contribution to the BI at \qbb\ prior
to active background reduction is listed in
\autoref{tab:parameter_estimates}. The statistical uncertainties on the
single contributions to the BI are generally of the order of 10\% or
lower, with the exception of \kvz\ and energy-degraded $\alpha$-events,
for which the uncertainty is roughly double. The two contributions are
affected by a higher uncertainty because they are not bound by screening
measurements.

The background event distribution in the \onbb\ analysis window can be
well approximated with a constant function (see \autoref{fig:roishape}).
With this assumption, the BIs extracted from data are
$16.4^{+1.7}_{-1.6} \cdot 10^{-3}$~\ctsper\ for \enrBEGe\ and
$15.4^{+1.8}_{-1.6} \cdot 10^{-3}$~\ctsper\ for the \enrCoax\ data set.
These values agree well with the background model description presented
in \autoref{sec:results}. The BIs prior to further analysis cuts and
before the upgrade of the \gerda\ experiment to \phasetwo\ can be found
in reference~\cite{Agostini2013a}. For the \enrCoax\ data set the BI
prior to the upgrade of $(18 \pm 2) \cdot 10^{-3}$~\ctsper\ is very
consistent with the values presented here. The BI of the \enrBEGe\ data
set instead is substantially improved from a \phaseone\ value of
$42^{+10}_{-8} \cdot 10^{-3}$~\ctsper\ to a value which is at least
$2.5\times$ smaller in \phasetwo\ despite a significant increase of
inactive hardware mass.\footnote{Note the slight difference of the
\enrBEGe\ analysis data set presented here and the data set used for
\onbb\ analysis for which the improvement in the BI is slightly higher
($3\times$ better BI). This is due to discarded $^\text{enr}$BEGe data
for which no PSD can be applied.} Contributions to the BI from all
isotopes have been improved with respect to \phaseone\ with the
exception of background introduced by $\alpha$ surface events. The most
drastic improvement is notable for \kvz\ for which the BI contribution
for the $^{\text{enr}}$BEGe detectors appears four times smaller than
before the upgrade to \phasetwo.

As mentioned in \autoref{sec:results}, the extracted \twonu\ half-life
estimate is based on the \enrBEGe\ data set only. The additional
parameter $\delta^{2\nu}$ mainly quantifies the systematic biases
between the active volume determination methods of the two detector
types. The full charge collection depth (FCCD), which determines the
active volume of a detector, was studied extensively in a detector
characterization campaign for the $^\text{enr}$BEGe
detectors~\cite{Agostini2015e,GERDAcollaboration2019}. The estimate of
the FCCD used in this analysis is based on measurements using an \Am\
source with characteristic $\gamma$-lines at 60~keV, 99~keV and 103~keV.
However, the FCCD was also measured using a \Co\ source with
characteristic $\gamma$ energies of 1173~keV and 1332~keV. The latter
FCCD$_\mathrm{Co}$ is systematically higher (about 3\%) with respect to
the FCCD$_\mathrm{Am}$. The discrepancy could be explained by an energy
dependence of the initial charge-carrier cloud size inside the detector
but the actual impact on the active volume is still under investigation.
For the $^\text{enr}$Coax detectors only FCCD values determined with a
\Co\ source are available. Considering the systematic uncertainties
affecting the determined active \gesix\ exposures of the \enrBEGe\ and
\enrCoax\ data sets (1.8\% and 5\% respectively, see
\autoref{tab:datasetdesc}) $\delta^{2\nu}$ is compatible with zero
within $1\sigma$.\footnote{The systematic bias between the active volume
estimates for the BEGe and coaxial detector types is a sub-dominant
contribution in the \onbb\ analysis with respect to e.g.~PSD
uncertainties.}

Various systematic effects have to be considered when estimating the
uncertainty on the \twonu\ half-life \thalftwo. Due to the fact that the
aim of the paper is not a precise \twonu\ half-life measurement, for
most of them only a conservative evaluation is provided. Several
systematic uncertainties arise from the Monte Carlo simulation
framework. Uncertainties due to the \geant\ model of particle
interactions and propagation were estimated to be of the order of 2\% in
previous publications~\cite{Agostini2013b,Agostini2015b}. Approximations
in the implementation of the \gerda\ setup are conservatively estimated
within a $1-2$\% uncertainty range. This accounts for possible spectral
shape modifications due to inaccurate charge collection model between
the \nplus\ contact layer and the active detector volume. Uncertainties
induced by the theoretical model of \twonu\ decays implemented in
\textsc{Decay0}, as well as data acquisition and selection methods are
considered negligible. A 1.8\% contribution accounts for uncertainties
in the enrichment and active mass fraction determination (see active
\gesix\ exposure in \autoref{tab:datasetdesc}). All the systematic
effects considered above sum up to a total systematic uncertainty on
\thalftwo\ of $3-4$\%. In total this leads to $T^{2\nu}_{1/2} = (2.03
\pm 0.09) \cdot 10^{21}$~yr compatible with earlier results
\cite{Agostini2013b,Agostini2015b}.


\section{Conclusions}%
\label{sec:conclusions}

We presented the background decomposition of \gerda\ \phasetwo\ data
before the application of active background suppression techniques using a multivariate
Bayesian fit approach based on single- and two-detector data in energy
and detector space. The model is able to well describe the data and the
results are compatible with the expectations from material screening
measurements. The only exception is \kvn\ for which a higher
contamination is found, dominantly in hardware components close to the
detector array. This indicates contaminations introduced during
production and mounting procedures different from the screened reference
samples; in fact a few parts underwent further processing after material
screening. Analyzing the count rates in the \kvn\ and \kvz\
high-statistics $\gamma$-lines on a detector-by-detector basis we find
indications for asymmetries in the spatial distribution of the two
potassium isotopes. Furthermore, the background indices at \qbb\ prior
active background suppression techniques are given by
\[
  \begin{split}
    \text{$^\text{enr}$BEGe} \quad &
    16.04^{+0.78}_{-0.85}\,\text{(stat)} \cdot
    10^{-3}\;\text{\ctsper} \\
    \text{$^\text{enr}$Coax} \quad &
    14.68^{+0.47}_{-0.52}\,\text{(stat)} \cdot
    10^{-3}\;\text{\ctsper} \\
  \end{split}
\]
and are in very good agreement with the assumption of a flat background
distribution in this region. In terms of the BI the upgrade to \gerda\
\phasetwo\ proves extremely successful. Despite major hardware changes
and higher inactive mass close to the detectors, the BI before applying
active background reduction remains unchanged for the $^\text{enr}$Coax
detectors and is improved by a factor of three for the $^\text{enr}$BEGe
detectors.

A careful background model is essential in order to separate the
two-neutrino double-beta decay events from the other background
components. We expect to substantially improve the precision of the
\thalftwo\ measurement after applying the LAr veto cut. In this manner,
the signal to background ratio in the \twonu\ energy region is improved
by about an order of magnitude~\cite{Agostini2017,Agostini2018}.
Furthermore, this allows precision studies of the shape of the \twonu\
spectrum and hence to test physics models beyond the Standard Model such
as \onbb\ decay with Majoron emission and Lorentz symmetry violation
effects~\cite{Agostini2015b,Diaz2014}.

The localization of impurities makes the exchange of particularly
contaminated components possible in upgrade works and thus the
background can be potentially lowered even further. Moreover, it is
important to learn what are the most important sources of background in
order to improve handling and cleaning procedures as well as material
selection. For future experiments like the Large Enriched Germanium
Experiment for Neutrinoless $\beta\beta$ Decay
(\textsc{Legend})~\cite{Abgrall2017}, which aims to (partly) cover the
parameter space of inverted neutrino mass hierarchy, background
reduction is the most crucial step in achieving the necessary
sensitivity. The goal is to achieve a background index one order of
magnitude lower than \gerda\ \phasetwo.


\section{Acknowledgments}%
\label{apdx:ack}
The \gerda\ experiment is supported financially by
  the German Federal Ministry for Education and Research (BMBF),
  the German Research Foundation (DFG) via the Excellence Cluster Universe
  and the SFB1258,
  the Italian Istituto Nazionale di Fisica Nucleare (INFN),
  the Max Planck Society (MPG),
  the Polish National Science Centre (NCN),
  the Foundation for Polish Science (TEAM/2016-2/2017),
  the Russian Foundation for Basic Research (RFBR), and
  the Swiss National Science Foundation (SNF).
The institutions acknowledge also internal financial support.
This project has received funding or support from the European Union’s
Horizon 2020 research and innovation programme under the Marie
Sklodowska-Curie Grant Agreements No. 690575 and No. 674896,
respectively.
The \gerda\ collaboration thanks the directors and the staff of the LNGS
for their continuous strong support of the \gerda\ experiment.

\clearpage
\appendix

\section{Potassium tracking analysis}%
\label{apdx:kmodel}

The two full-energy lines of \kvn\ and \kvz\ at 1461~keV and 1525~keV
are distinct features of the energy spectrum shown in
\autoref{fig:datadesc}. Being a relevant source of background for
double-beta decay, the two potassium isotopes play a crucial role in the
background modeling process in \gerda. Uncertainties in their origin and
distribution propagate directly to searches for exotic physics like
Majorons, Lorentz invariance-violating processes or decay modes to
excited states of \twonu\ decay in which the shape of the \twonu\ decay
spectrum is a unique feature and thus need to be well understood. 

Initial observations in \phasetwo\ have shown that the \kvn\ and \kvz\
full-energy line intensities have increased by a factor of 4 and 2,
respectively, in the single-detector data compared to
\phaseone~\cite{dandrea2017}. The \kvz\ increase in activity can be
attributed to the exchange of the mini-shrouds material from copper to
nylon during the \phasetwo\ upgrade: The electric field generated by the
detectors bias high voltage is not screened by the conductive material
anymore. The \kvz\ ions can be attracted from a larger LAr volume
into the vicinity of the detectors. Moreover, the unshielded
high-voltage cables could be an explanation for the higher rate of \kvz\
events seen in the uppermost detectors in the \gerda\ array. The higher
\kvn\ event rate, on the other hand, is possibly attributable to the
glue used for the nylon mini-shrouds and other new materials introduced
with the LAr veto system. The exact amount, location and radio-purity of
the glue is not precisely known.

In the following sections we focus on the characteristics of the events
constituting the two potassium lines. In order to extract information
about the spatial distribution of \kvn\ and \kvz\ contamination around
the \gerda\ array, a treatment on a detector-by-detector basis is
advantageous. The two $\gamma$-lines contain enough statistics for such
an analysis to be meaningful and constitute samples with a high signal
to background ratio.

\subsection{Data}\label{apdx:kmodel:data}

Two windows around the potassium $\gamma$-lines are projected in
detector index space, such that, for single-detector data, each data
point $n_i$ represents the total counts in detector $i$ in the
respective energy window. For two-detector data the detector space is
two-dimensional, and each data point $n_{ij}$ represents the number of
events for which energy is deposited in detector $i$ and detector $j$.

\begin{figure}[tbp]
  \centering
  \includegraphics[width=0.8\linewidth]{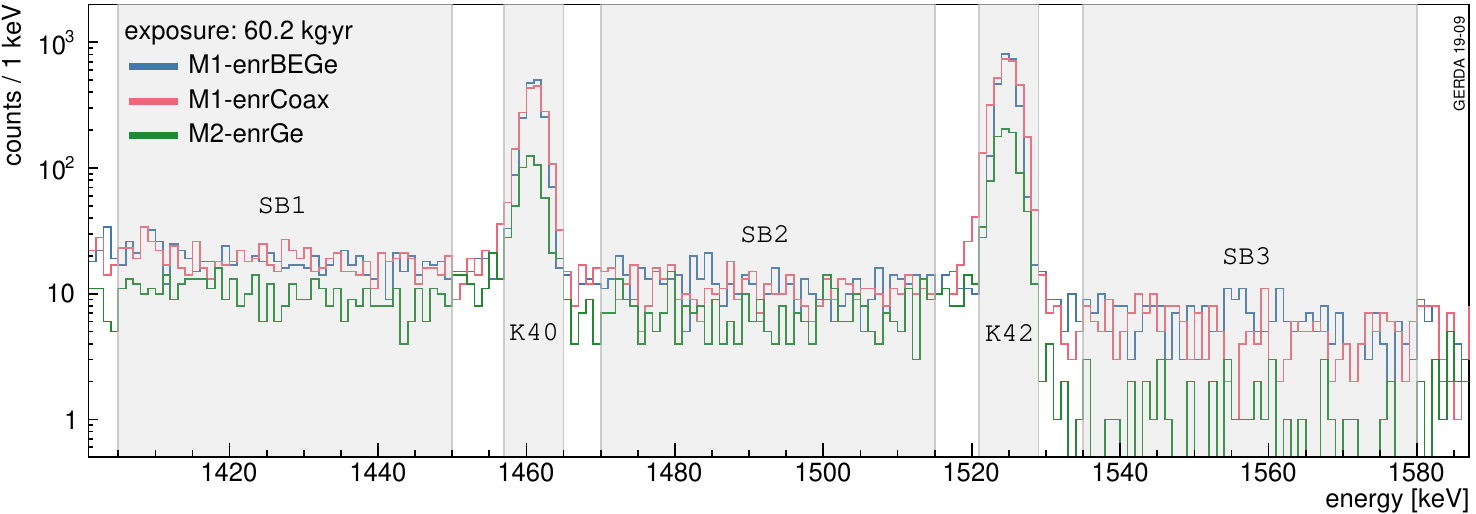}
  \caption{%
    Visual representation of the five energy ranges defined for the
    potassium tracking analysis. The exact intervals and counts are
    given in \autoref{tab:kmodel:k-regions-cts}.
  }\label{fig:k-regions}
\end{figure}

The events in the potassium lines (denoted with \m{K40} and \m{K42} in
the following) are selected in a $\pm3\sigma$ energy interval around the
respective line, rounded up to an integer number of keV to match the
specific energy windows in the energy distributions with 1~keV binning.
$\sigma$ is the energy resolution in the respective energy window.
Additionally, three side-bands (\m{SB1}, \m{SB2} and \m{SB3} in the
following) are used to estimate the continuum below and above the
$\gamma$-lines. Considering the further subdivision in single- (\m{M1-})
and two-detector (\m{M2-}) data, this leads to the definition of $5
\times 2$ energy regions, summarized
in~\autoref{tab:kmodel:k-regions-cts}. A visual representation of the
selected windows can be found in~\autoref{fig:k-regions}. We use the
PDFs respective to \Bih\ on the flat cables and detector intrinsic
\nnbb\ decays to estimate the background. Other components are expected
to contribute less in the respective energy windows.

\begin{table}[ht]
  \centering
  \caption{%
    Energy ranges and corresponding number of events for the potassium
    tracking analysis (visualized in \autoref{fig:k-regions}). Note that
    the windows for two-detector data are larger as the two
    single-detector energy resolutions are folded in the summed energy
    spectrum.
  }\label{tab:kmodel:k-regions-cts}
  \begin{tabular}{ccccc}
    \toprule
                  & \m{M1-} [keV] & cts. & \m{M2-} [keV] & cts. \\
    \midrule
    \m{K40}       & $[1457,1465]$ & 4472 & $[1455,1467]$ & 554  \\
    \m{K42}       & $[1521,1529]$ & 6718 & $[1519,1531]$ & 865  \\
    \midrule
    \m{SB1}       & $[1405,1450]$ & 1852 & $[1405,1450]$ & 452  \\
    \m{SB2}       & $[1470,1515]$ & 1124 & $[1470,1515]$ & 326  \\
    \m{SB3}       & $[1535,1580]$ & 533  & $[1535,1580]$ & 41   \\
    \bottomrule
  \end{tabular}
\end{table}

\subsection{Analysis}%
\label{apdx:kmodel:ana}

The statistical approach of factorizing the likelihood is described in
\autoref{subsec:likelihoodfact}. The part of the likelihood we are
analyzing here runs simultaneously on the $5 \times 2$ energy ranges
presented above. Following the naming convention introduced
in~\autoref{sec:stat-ana} it reads:
\[
  \mathcal{L}_\text{K}(\lambda_1,\ldots,\lambda_{m'}\,|\,n) =
  \prod_{d=1}^{N_\text{dat}}
  \left\{
    \prod_{i=1}^{N_\text{det}}
    \text{Pois}(n_{d,i}^\text{\m{M1}};\nu_{d,i}^\text{\m{M1}})
    \times
    \prod_{j<k}^{N_\text{det}}
    \text{Pois}(n_{d,jk}^\text{\m{M2}};\nu_{d,jk}^\text{\m{M2}})
  \right\}\;,
\]
where the index $i$ runs over the bins (i.e.~detectors) and the index
$d$ over the 5 considered energy windows, namely the three side-bands
\m{SB1}, \m{SB2}, \m{SB3} and the two line-bands \m{K40} and \m{K42}.
The \m{M2-} data sets are two-dimensional in detector space and run over
the two indices $j$ and $k$.

Gaussian prior probability distributions for the \kvn\ activity are
built from radio-purity screening measurements (see
reference~\cite{Agostini2018a} Sec.~5). For \kvz, for which no screening
information is available, uniform priors are adopted, with the exception
of the two \kvz\ components located on the \nplus\ contact surface of
$^\text{enr}$BEGe and $^\text{enr}$Coax detectors. \kvz\ can be
attracted to the \nplus\ surface by the electrical field created by the
high voltage potential applied to the detectors. Both components are
expected to be correlated by the volume ratio of the mini-shrouds (3:2
$^\text{enr}$BEGe to $^\text{enr}$Coax) the \kvz\ ions are attracted
from. The volume ratio estimate is extracted from the geometric
implementation in \mage. We assume an uncertainty of 0.1~mBq on either
activity allowing for a change of their ratio. The correlation is
included in the fit via a two-dimensional prior. 

The analysis flow starts with a construction of a first, preliminary
model, which consists only of background contributions that are expected
from screening measurements of \kvn\ and known properties of \kvz.
The resulting model, however, gives a non-satisfactory description of
data and the posterior distributions for the \kvn\ components are
significantly shifted to higher values with respect to the prior
distributions, indicating a surplus of \kvn.


To find a better agreement with physics data while keeping the model as
simple as possible, additional components using uniform priors are
included one at a time in the fitting procedure, and the Bayes factor is
calculated between the extended and the preliminary model. The model is
iteratively updated by adding the component that results in the highest
Bayes factor until no Bayes factor is larger than 10.

In a first iteration a replica of the PDF of \kvn\ in the mini-shrouds
is added obtaining a Bayes factor $\gg10$. \kvn\ in the \tetratex-coated
copper shrouds is added in a second iteration with a Bayes factor of 11.
For \kvz\ the only additional component that results in a Bayes factor
greater than 1 is \kvz\ on the \nplus\ detector contacts. Although the
fit shows only a slight preference (Bayes factor of 2) the component is
added to the model because of its importance in the full-range fit,
where the energy region above the 1525~keV $\gamma$-line is also
considered.

\begin{figure}[tbp]
  \centering
  \includegraphics[width=0.45\linewidth]{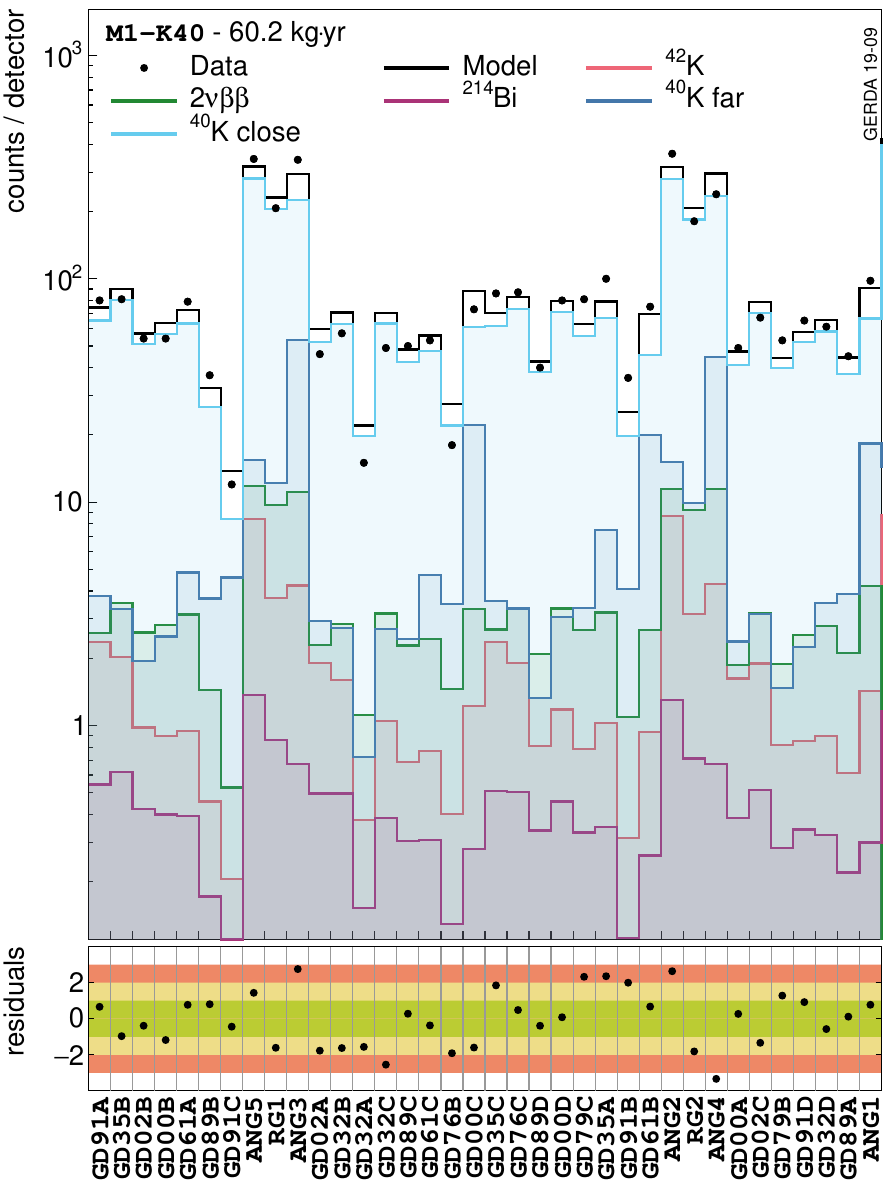}
  \includegraphics[width=0.45\linewidth]{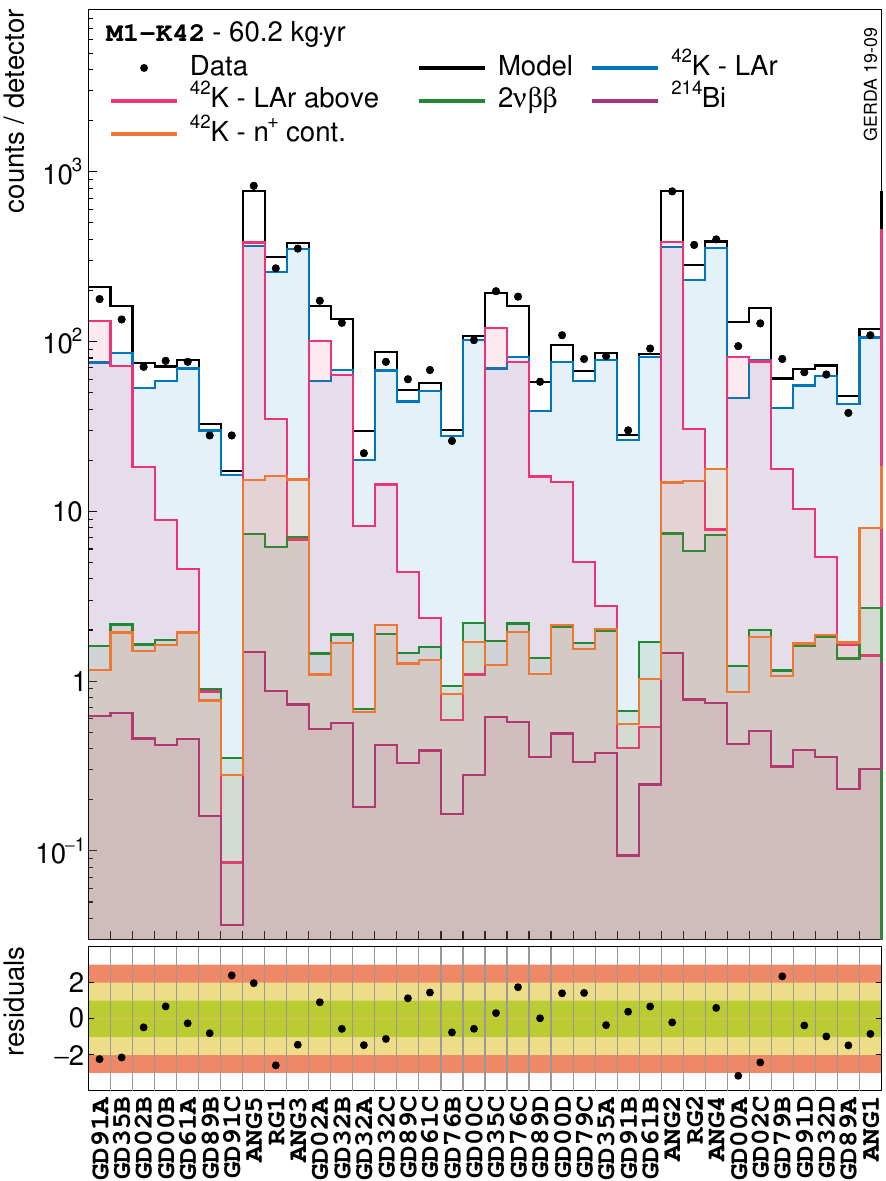}
  \includegraphics[width=0.45\linewidth]{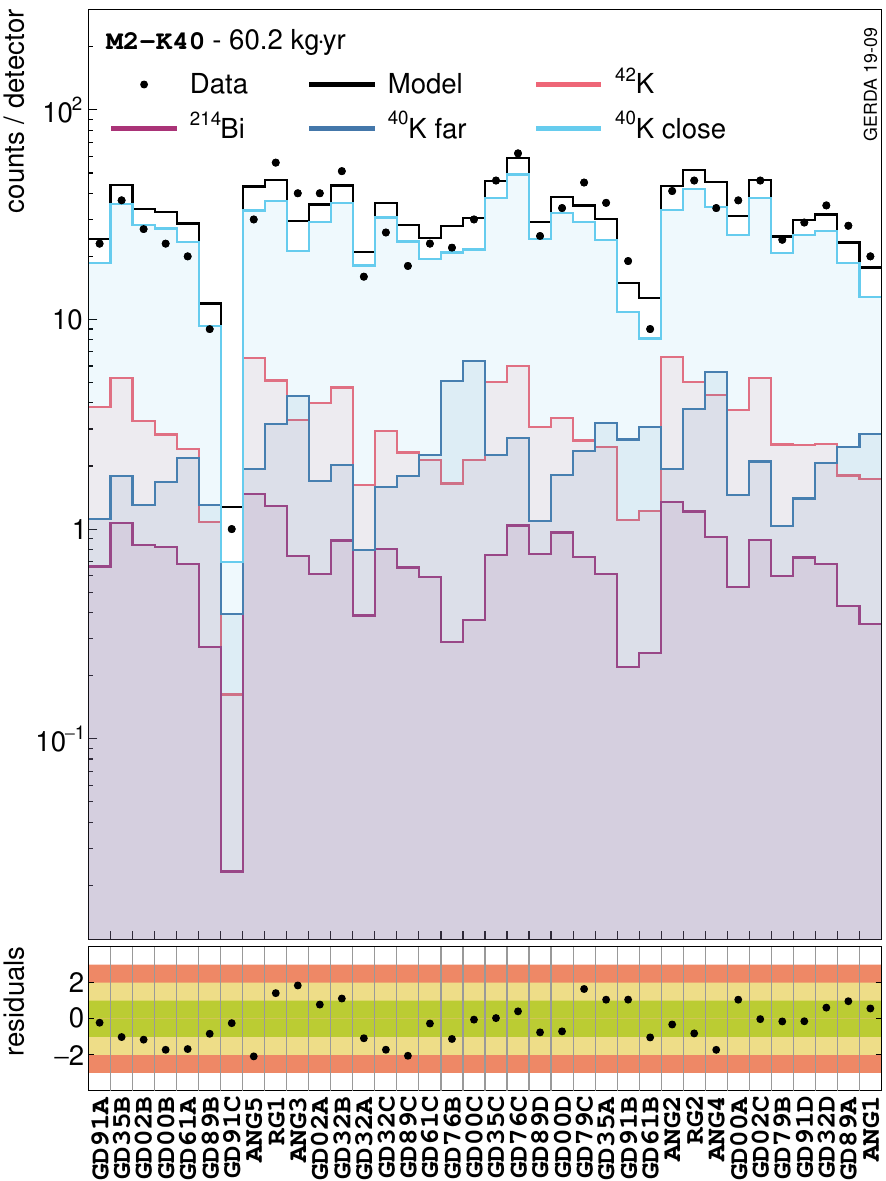}
  \includegraphics[width=0.45\linewidth]{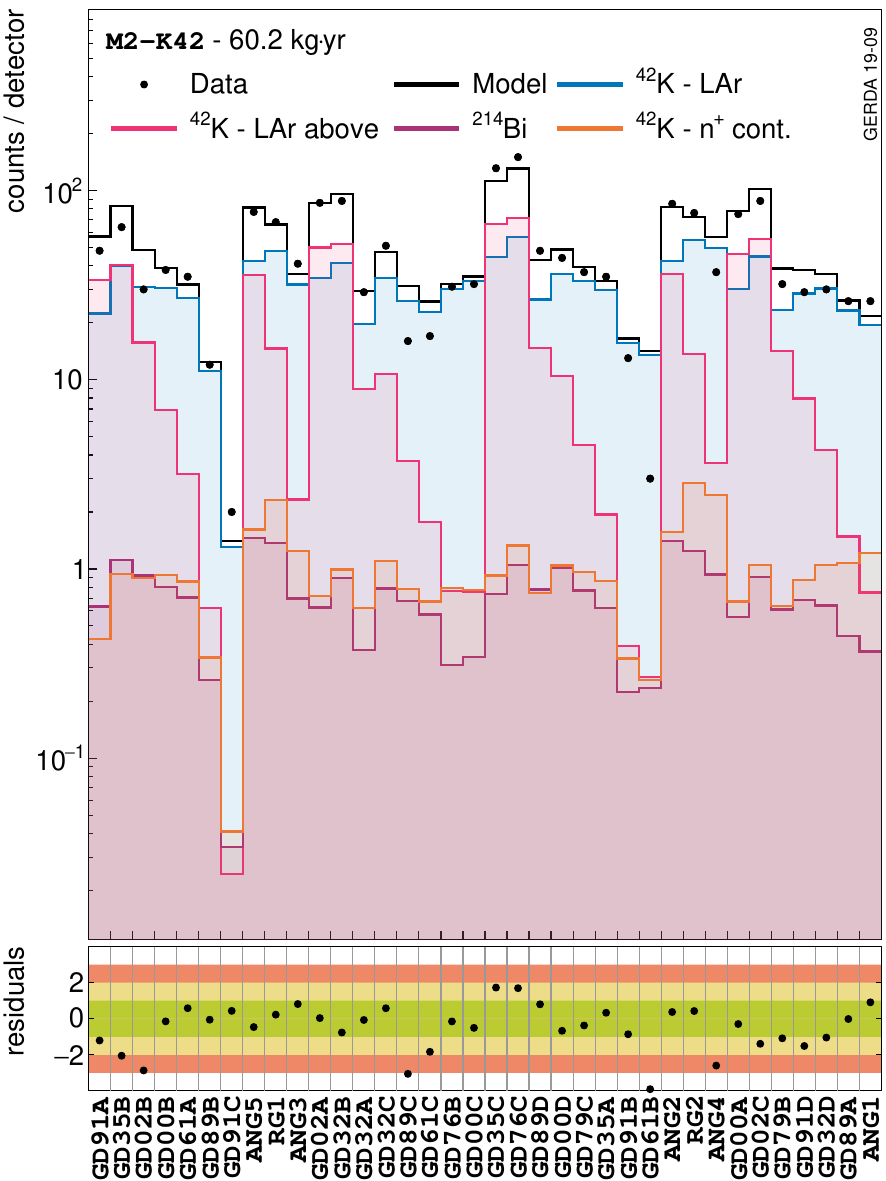}
  \caption{%
    Decomposition of the energy windows corresponding to the two
    potassium lines in detector space: single-detector data (top)
    one-dimensional representation of two-detector data (bottom). Some
    components are merged for visualization purposes: in the \m{K40}
    plots combined components are shown for \kvz\ and \Bih, while \kvn\
    sources are grouped in close (flat cables, holders, mini-shrouds)
    and far (fibers, SiPMs, copper shrouds, front-end electronics)
    locations from the detector array. To visualize the two-detector
    data the sum of the projections on the two domain axes (index $i$
    and index $j$) is shown.
  }\label{fig:kmodel:spc}
\end{figure}

The results of the base model are shown in \autoref{tab:kmodel:base} and
a graphic representation showing the counts per detector in 
both potassium $\gamma$-lines in \m{M1-} and \m{M2-}data can be found in
\autoref{fig:kmodel:spc}. The analysis yields a \pvalue\ of $\sim0.07$,
indicating an acceptable description of the data. To further improve the
model rotationally asymmetric fit components are needed. The base model
is accurate enough to be used as input for the full-range fit, which is
insensitive to any rotational inhomogeneity of the location of
background sources, as spectra from different detectors are merged into
a single data set.

\begin{table}[tbp]
  \centering
  \caption{%
    Summary of the fit parameters estimated with the potassium source
    tracking analysis (base model). The type of prior distribution is
    indicated with \m{[f]}: flat, \m{[g]}: Gaussian. ($\,^{\dagger}$
    \tetratex-coated)
  }\label{tab:kmodel:base}

\begin{tabular}{rlcccc}
  \toprule
  \multirow{2}{*}{source}        & \multirow{2}{*}{\m{[prior]} location} & \multirow{2}{*}{units} & global & marg. & 68\% C.I. or          \\
                                 &                                       &                        & mode   & mode  & 90\% upper C.L. \\
  \midrule
  \multirow{9}{*}{\kvn}          & \m{[g]} flat cables                   & \multirow{9}{*}{mBq}   & 3.29   & 3.25  & $[1.79, 4.72]$      \\
                                 & \m{[g]} front-end electronics         &                        & 15.7   & 15.9  & $[11.1, 20.1]$      \\
                                 & \m{[g]} copper shrouds $^{\dagger}$   &                        & 18.4   & 18.1  & $[16.6, 20.0]$      \\
                                 & \m{[g]} fiber shroud                  &                        & 2.82   & 2.81  & $[2.24, 3.38]$      \\
                                 & \m{[g]} detector holders              &                        & 1.73   & 1.73  & $[1.28, 2.14]$      \\
                                 & \m{[g]} mini-shrouds                  &                        & 1.70   & 1.70  & $[1.60, 1.80]$      \\
                                 & \m{[g]} SiPM ring                     &                        & 2.50   & 2.73  & $[0.83, 4.13]$      \\
                                 & \m{[f]} far from the array            &                        & 328    & 322   & $[232,  416]$       \\
                                 & \m{[f]} close to the array            &                        & 10.8   & 10.8  & $[9.53, 12.1]$      \\
  \midrule
  \multirow{4}{*}{\kvz}          & \m{[f]} \nplus\ (BEGe)                &  \multirow{4}{*}{mBq}  & 0      & 0     & < 0.37              \\
                                 & \m{[f]} \nplus\ (Coax)                &                        & 0.22   & 0.24  & $[0.12, 0.38]$      \\
                                 & \m{[f]} LAr -- above array            &                        & 450    & 454   & $[436,  470]$       \\
                                 & \m{[f]} LAr -- outside mini-shrouds   &                        & 2036   & 2009  & $[1915, 2080]$      \\
  \midrule
  \Bih                           & \m{[g]} flat cables                   &                mBq     & 1.51   & 1.26  & $[0.93, 1.51]$      \\
  \midrule
  \nnbb\                         & \m{[f]} germanium                     & $10^{21}$yr            & 1.91   & 1.93  & $[1.86, 2.00]$      \\
  \bottomrule
\end{tabular}


\end{table}

\begin{figure}[tpb] \centering
  \includegraphics[width=0.6\linewidth]{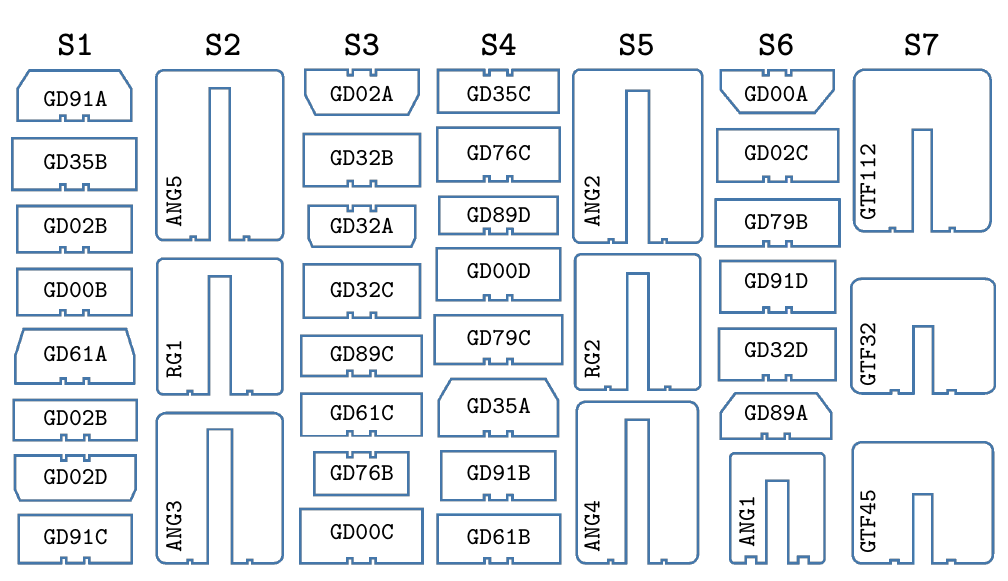}
  \caption{Detector string configuration in the \gerda\ array. Names
    prefixed with \texttt{GD} refer to detectors of $^\text{enr}$BEGe type
    whereas \texttt{ANG} and \texttt{RG} refer to $^\text{enr}$Coax
    detectors. The three natural coaxial detectors (prefixed with
    \texttt{GTF}) which are located in the central string \texttt{S7}
    are not used in this analysis.
  }\label{fig:detstrings}
\end{figure}


The two components \textit{\kvn\ close to the array} and \textit{\kvz\
in LAr -- above the array} are split into 7 sub-components on a
string-by-string basis (for the respective PDFs see
\autoref{apdx:pdfs}). Furthermore, we consider a \kvn\ contamination on
top of the central mini-shroud.

The results of this extended analysis are listed in
\autoref{tab:kmodel:ext}. An elevated \kvz\ concentration is found above
the central string while a lower concentration is observed above the
adjacent strings \m{S1} and \m{S6} (string numbers follow the
nomenclature used in \autoref{fig:detstrings}). Due to the large number
of components the fit yields a high anti-correlation between the \kvz\
concentration above the outer strings and \m{S7}. This results in a high
uncertainty on the latter fit parameter. 

The screening measurements do not account for all observed \kvn. In
general ICP-MS screening of the mini-shrouds with respect to \kvn\ is
difficult and yielded only a lower limit. Different measurements seem to
indicate different contamination levels of different mini-shrouds.
Samples of glued nylon yielded the highest potassium contamination. As
the gluing of the nylon mini-shrodus is done manually during
installation the amount of glue and its exact location is hard to
control. Hence, an asymmetric distribution is expected. The \kvn\
content of other close components like holders and cables might also be
asymmetric. The asymmetric \kvn\ contamination is confirmed by the
extended potassium tracking analysis. Also, an additional \kvn\
distribution on the top-lid of the central mini-shroud is preferred.
The surplus far \kvn\ component instead is possibly explained by setup
parts omitted in the model like the PMTs and voltage-dividers of the LAr
veto system. An upper limit of their \kvn\ content, $<330$~mBq, was
estimated from material screening which is similar to the activity
reconstructed for the far \kvn\ component. The location of the PMTs with
respect to the detector array is very similar to the \tetratex-coated
copper-shrouds and their PDFs are, hence, degenerate.

\begin{table}[tbp]
  \centering
  \caption{%
    Summary of the fit parameters estimated with the potassium source
    tracking analysis (extended model). The type of prior distribution
    is indicated with \m{[f]}: flat, \m{[g]}: Gaussian. ($\,^{\dagger}$
    \tetratex-coated)
  }\label{tab:kmodel:ext}

\begin{tabular}{rlcccc}
  \toprule
  \multirow{2}{*}{source}    & \multirow{2}{*}{\m{[prior]} location} & \multirow{2}{*}{units} & global & marg. & 68\% C.I. or          \\
                             &                                       &                        & mode   & mode  & 90\% upper C.L. \\
  \midrule
  \multirow{16}{*}{\kvn}     & \m{[g]} flat cables                   & \multirow{16}{*}{mBq}  & 2.33   & 1.08  & $[0.13,2.30]$       \\
                             & \m{[g]} front-end electronics         &                        & 14.5   & 14.4  & $[10.2,18.7]$       \\
                             & \m{[g]} copper shrouds $^{\dagger}$   &                        & 18.4   & 18.5  & $[16.6,20.0]$       \\
                             & \m{[g]} fiber shroud                  &                        & 2.83   & 2.77  & $[2.24,3.38]$       \\
                             & \m{[g]} detector holders              &                        & 2.57   & 2.29  & $[1.75,2.78]$       \\
                             & \m{[g]} mini-shrouds                  &                        & 1.70   & 1.70  & $[1.60,1.79]$       \\
                             & \m{[f]} close to \m{S1}               &                        & 0.81   & 0.83  & $[0.47,1.28]$       \\
                             & \m{[f]} close to \m{S2}               &                        & 2.35   & 2.22  & $[1.83,2.51]$       \\
                             & \m{[f]} close to \m{S3}               &                        & 0      & 0     & < 0.50              \\
                             & \m{[f]} close to \m{S4}               &                        & 2.58   & 2.55  & $[2.10,3.02]$       \\
                             & \m{[f]} close to \m{S5}               &                        & 0.97   & 0.85  & $[0.56,1.16]$       \\
                             & \m{[f]} close to \m{S6}               &                        & 1.86   & 1.89  & $[1.46,2.30]$       \\
                             & \m{[f]} close to \m{S7}               &                        & 0      & 0     & < 2.92              \\
                             & \m{[f]} \m{S7} mini-shroud (top)      &                        & 2.09   & 1.83  & $[1.26,2.40]$       \\
                             & \m{[g]} SiPM ring                     &                        & 2.44   & 2.32  & $[0.83,4.02]$       \\
                             & \m{[f]} far from the array            &                        & 390    & 374   & $[280,468]$         \\
  \midrule
  \multirow{10}{*}{\kvz}     & \m{[f]} \nplus\ (BEGe)                &  \multirow{10}{*}{mBq} & 0.15   & 0.19  & $[0.05,0.37]$       \\
                             & \m{[f]} \nplus\ (Coax)                &                        & 0.22   & 0.26  & $[0.12,0.41]$       \\
                             & \m{[f]} LAr -- above \m{S1}           &                        & 0      & 0     & < 0.80              \\
                             & \m{[f]} LAr -- above \m{S2}           &                        & 2.22   & 2.96  & $[2.21,3.63]$       \\
                             & \m{[f]} LAr -- above \m{S3}           &                        & 1.20   & 1.57  & $[1.06,2.16]$       \\
                             & \m{[f]} LAr -- above \m{S4}           &                        & 1.43   & 1.89  & $[1.33,2.41]$       \\
                             & \m{[f]} LAr -- above \m{S5}           &                        & 1.49   & 1.91  & $[1.38,2.73]$       \\
                             & \m{[f]} LAr -- above \m{S6}           &                        & 0      & 0     & < 1.21              \\
                             & \m{[f]} LAr -- above \m{S7}           &                        & 10.4   & 7.84  & $[4.95,9.83]$       \\
                             & \m{[f]} LAr -- outside mini-shrouds   &                        & 2083   & 2058  & $[1960,2145]$       \\
  \midrule
  \Bih                       & \m{[g]} flat cables                   &                 mBq    & 1.60   & 1.41  & $[1.14,1.66]$       \\
  \midrule
  \nnbb\                     & \m{[f]} germanium                     & $10^{21}$yr            & 1.89   & 1.89  & $[1.83,1.97]$       \\
  \bottomrule
\end{tabular}

\end{table}



\clearpage
\section{\texorpdfstring{$\alpha$-events background analysis}{a-events background analysis}}%
\label{apdx:amodel}

Above an energy of 3.5~MeV almost all registered events are due to
$\alpha$-emitting isotopes. The respective part of the full likelihood
can be approximately factorized and studied separately.
$\alpha$-particles have a very short range in LAr as well as in
germanium (continuous slowing down approximation, CSDA, range of
$50~\upmu$m and $20~\upmu$m, respectively~\cite{astar}) and are able to
reach a detector's active volume only through the very thin (of the
order of $500$~nm) \pplus\ contact surface. Therefore, the
$\alpha$-emitter contamination is detector-specific and depends only on
the \pplus\ surface contaminations. Therefore, we analyze the
$^\text{enr}$BEGe and $^\text{enr}$Coax detector data separately in
energy space; the projection in detector space bares no correlation
between detectors and hence contains no further useful information. The
number of events in a single detector is not sufficient to further split
the data on a detector-by-detector basis. The two data sets are
uncorrelated and the statistical analysis can be carried out for each
single-detector data set separately. In the two-detector data the
$\alpha$-component is not observed due to the short range of these
particles.

All contaminations found are constituents of the \Uh\ decay chain. The
main surface contamination observed is \Po\ which occurs either as an
incident contamination and decays in time with a half-life of
$138.3763(17)$~days~\cite{Be2008} or is fed by a contamination with
\Pbl\ with a stable rate in time. The spectral form is identical for
both cases and can only be disentangled by analyzing the $\alpha$-rate
in time (see \autoref{subsec:timealpha}).

Above the \Po\ peak very few events are observed. In the \enrBEGe\ data
set we find only four events with an energy larger than 5.3~MeV, while
in the \enrCoax\ data set 22 such events are observed, 14 of which in a
single detector \texttt{ANG2} (see~\autoref{tab:AperDet}). These events
are due to $\alpha$-decays from \Rn\ and subsequent isotopes on the
\pplus\ detector surfaces. \texttt{ANG2} also shows a higher \radzzs\
(mother nucleus of \Rn) contamination which suggests dominantly a
surface contamination with \radzzs\ rather than \Rn\ dissolved in the
LAr. In the latter case the decay chain would be broken as only the
gaseous \Rn\ emanates from the construction materials into the LAr.
Hence, in the following, we will only consider a \pplus\ surface
contamination with \radzzs\ and all subsequent isotopes to which we
refer as the \radzzs\ decay chain. The \Po\ and \radzzs\ contaminations
are not necessarily spatially correlated.

Due to the very short range of $\alpha$-particles the energy spectrum of
$\alpha$-decays exhibits a line with a pronounced low-energy tail. The
tail is formed when the decay occurs under an incident angle with
respect to the contact and the $\alpha$-particle loses part of its
energy before reaching the detectors active volume. The maximum is
shifted with respect to the full emission energy which is due to energy
loss inside the electrode and depends on its minimal thickness. The
detectors have slightly different contact thicknesses, also, the \pplus\
contact of a single detector may intrinsically be inhomogeneous.
Therefore, we model the \Po\ peak with a mixture of PDFs obtained from
simulations with different contact thicknesses. Due to the low number of
counts observed in the \radzzs\ chain it is sufficient to model this
component with only one PDF. Furthermore, the isotope contamination is
assumed to halve at each decay step. A reduction effect of the
subsequent $\alpha$-decays in the \Rn\ chain had been observed in
\phaseone\ and attributed to possible recoil off the surface into the
LAr~\cite{Agostini2014}. We adopt this explanation in our model although
we note that the number of events observed with an energy >5.3~MeV is
not sufficient to confirm the previously observed reduction effect.
Further details about the construction of the PDFs are given in
\autoref{apdx:pdfs}.

Dedicated measurements~\cite{Agostini2013thesis} have shown that events
originating in the contact separating groove are partly reconstructed
with degraded energy. A simulation-based model of these energy-degraded
events is not available yet. We approximate this component with an
empirical linear distribution truncated below the maximum of the \Po\
peak. Such a component accommodates also eventual $\alpha$-decays in the
LAr in very close vicinity to the \pplus\ detector surface. However, the
number of events found with an energy >5.3~MeV is too low to fully
account for the linearly modeled distribution.

The likelihood function for modeling the high-energy region dominated by
$\alpha$-decays runs only on single-detector data, namely \enrBEGe\ and
\enrCoax\ separately, in a range from 3.5~MeV to 5.25~MeV. Events with
an energy higher than 5.25~MeV are put in a single overflow bin:
\begin{equation}
  \mathcal{L}_\alpha(\lambda_1,\ldots,\lambda_m\,|\,n) =
  \prod_{i=1}^{N_\text{bins}} \text{Pois}(n_{i};\nu_{i})\;
  \label{eq:apdx:alpha:likelihood}
\end{equation}
A flat prior probability is assigned to each of the fit parameters
$\lambda_i$. Both data sets are fit separately with a fixed bin size of
10~keV\footnote{The calibration curves are accurate on the sub-keV level
up to the highest $\gamma$-energy of about 2.6~MeV emitted by the \Th\
calibration sources. Although no major non-linearity effects were found
the same accuracy can not be guaranteed at 6~MeV. Deviations from
linearity at this energy are within 10~keV, hence, we increase the bin
size in the higher energy range.} as the $\alpha$-contamination is
detector individual and the two single-detector data sets are
uncorrelated in the respective energy window. 

\begin{table}[tb]
  \centering
  \caption{Observed number of counts with energy $>5.3$~MeV belonging to
    the \radzzs\ decay chain. Detectors with zero counts are not listed.
  }\label{tab:AperDet}
  \begin{tabular}{cccc}
    \toprule
    data set & detector & channel & \radzzs-chain [cts] \\
    \midrule
    \multirow{3}{*}{\texttt{M1-enrBEGe}} & \m{GD61C} & 16 & 1 \\
                                         & \m{GD79B} & 32 & 1 \\
                                         & \m{GD89A} & 35 & 2 \\
                                         \midrule
    \multirow{6}{*}{\texttt{M1-enrCoax}} & \m{ANG1} & 36 & 2 \\
                                         & \m{ANG2} & 27 & 14 \\
                                         & \m{ANG3} & 10 & 1 \\
                                         & \m{ANG4} & 29 & 1 \\
                                         & \m{ANG5} &  8 & 2 \\
                                         & \m{RG1}  &  9 & 2 \\
                                         \bottomrule
  \end{tabular}
\end{table}

\begin{figure}[tb]
  \centering
  \includegraphics{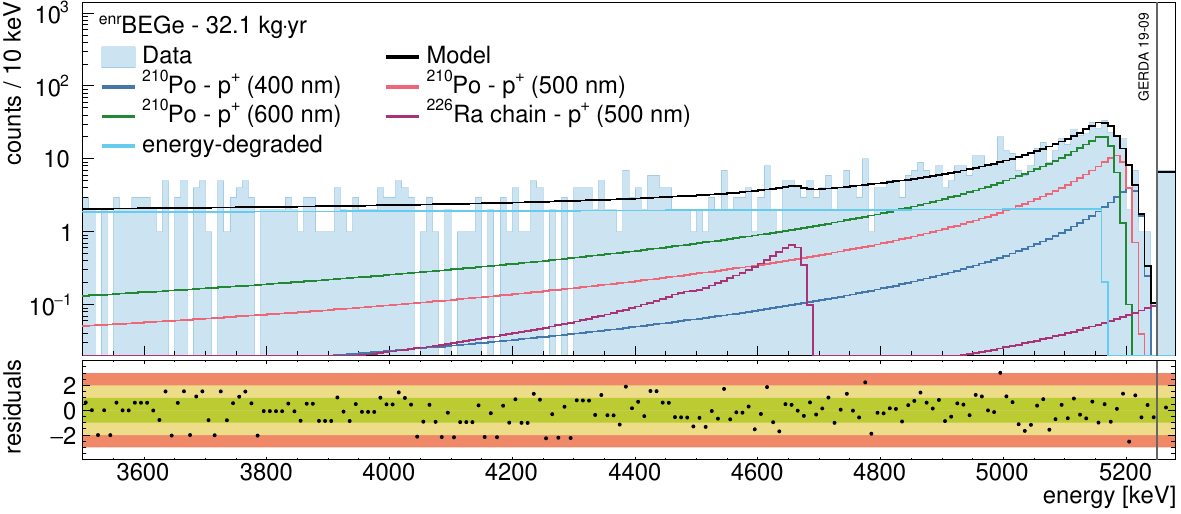}
  \includegraphics{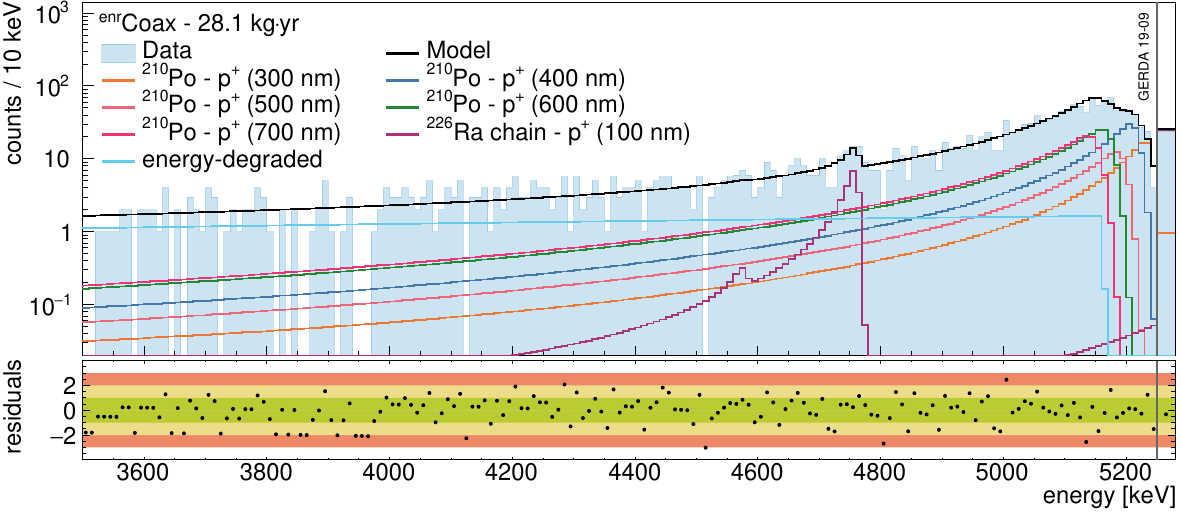}
  \caption{%
    Fit results of the $\alpha$-events background analysis for \enrBEGe\
    (top) and \enrCoax (bottom). The last bin contains all events above
    5250~keV.
  }\label{fig:apdx:alphafit}
\end{figure}

The fit results are shown in \autoref{fig:apdx:alphafit} and listed in
\autoref{tab:apdx:res}. The \Po\ component is modeled with a combination
of \pplus\ contact thicknesses from 400 to 600~nm for the \enrBEGe\ data
set and from 300 to 700~nm for the \enrCoax\ data set in steps of
100~nm. Further \Po\ components are rejected by a Bayes factor analysis.
Impurities belonging to the \radzzs\ chain are mostly located on
\texttt{ANG2} and thus a fit of the \enrCoax\ data set using a single
\pplus\ thickness describes this component well. For the \enrBEGe\ data
set we observe a very small number of counts for the \radzzs\ chain,
therefore, also in this case a single component is sufficient. We
determine a best-fit value of 100~nm and 500~nm, respectively. The
estimated \pvalue\ for \enrBEGe\ is 0.2 whereas the \pvalue\ for
\enrCoax\ is 0.3. The dominant spectral component below 4.5~MeV is due
to degraded $\alpha$-events which extends down to the ROI.

\begin{table}[tb]
\centering
  \caption{%
    Fit results of the $\alpha$-events background analysis for the
    \enrBEGe\ and \enrCoax\ data sets. Values are given in counts in the
    full PDF range from 40~keV to 8000~keV.
  }\label{tab:apdx:res}
  \begin{tabular}{rlccr@{ }l}
\toprule
\multirow{2}{*}{data set}      & \multirow{2}{*}{component} & contact & global mode & \multicolumn{2}{c}{marg.~mode}      \\
                               &                            & [nm]    & [cts]       & \multicolumn{2}{c}{68\% C.I.~[cts]} \\
\midrule
\multirow{6}{*}{\enrBEGe}      & \multirow{4}{*}{\Po}       & 400     & 49          & 50                                     & $[34,76]$      \\
                               &                            & 500     & 162         & 165                                    & $[107,222]$    \\
                               &                            & 600     & 346         & 342                                    & $[278,391]$    \\
                               &                            & comb.   & --          & 555                                    & $[523,586]$    \\
                               & \radzzs\ chain             & 500     & 20          & 20                                     & $[15,29]$      \\
                               & energy-degraded            & --      & --          & 845                                    & $[698,948]$    \\
\midrule
\multirow{8}{*}{\enrCoax}      & \multirow{6}{*}{\Po}       & 300     & 167         & 165                                    & $[140,208]$   \\
                               &                            & 400     & 363         & 368                                    & $[272,430]$   \\
                               &                            & 500     & 182         & 175                                    & $[83,338]$    \\
                               &                            & 600     & 433         & 420                                    & $[233,582]$   \\
                               &                            & 700     & 404         & 410                                    & $[295,537]$   \\
                               &                            & comb.   & --          & 1555                                   & $[1511,1609]$ \\
                               & \radzzs\ chain             & 100     & 58          & 59                                     & $[49,70]$     \\
                               & energy-degraded            & --      & --          & 485                                    & $[426,599]$   \\
\bottomrule
\end{tabular}


\end{table}

\subsection{\texorpdfstring{Time distribution of
$\alpha$-events}{Distribution in time of alpha-events}}%
\label{subsec:timealpha}

The time distribution of \Po\ decays is well known to be exponential,
however, in the presence of a \Pbl\ contamination a constant
contribution can also be observed. \Pbl, decaying to \Po, feeds a
constant \Po\ component once their decay rates stabilize in a secular
equilibrium. To disentangle the two we fit the time distribution of
events with energies between 3.5~MeV and 5.25~MeV with a constant $C$
and an exponential function:
\[
  f(t) = C + N \exp\left( - \frac{\log2}{T_{1/2}}t \right)
\]
where $T_{1/2}=(138.4\pm0.2)$~days is the half-life of \Po. We use a
Poisson likelihood function corrected for data acquisition dead
time~\cite{Cleveland1983} and model the time bin content as follows
\[
  \nu_i = f_i^{\mathrm{LT}}
  \left\{ C \delta t + N \tau
    \left[
      \exp\left( -\frac{t_0 + i \delta t}{\tau} \right)  -
      \exp\left( -\frac{t_0 + (i+1) \delta t}{\tau} \right)
    \right]
  \right\}
\]
$C$ and $N$ are the amplitudes of the constant and the exponentially
decaying components and are the only two free fit parameters.
$f_i^{\mathrm{LT}}$ is the live-time fraction in time-bin $i$ which is
estimated from injected test pulser events, $\delta t$ is the time-bin
width and $\tau = T_{1/2} / \log2$.

\begin{figure}[btp]
  \centering
  \includegraphics[width=0.8\textwidth]{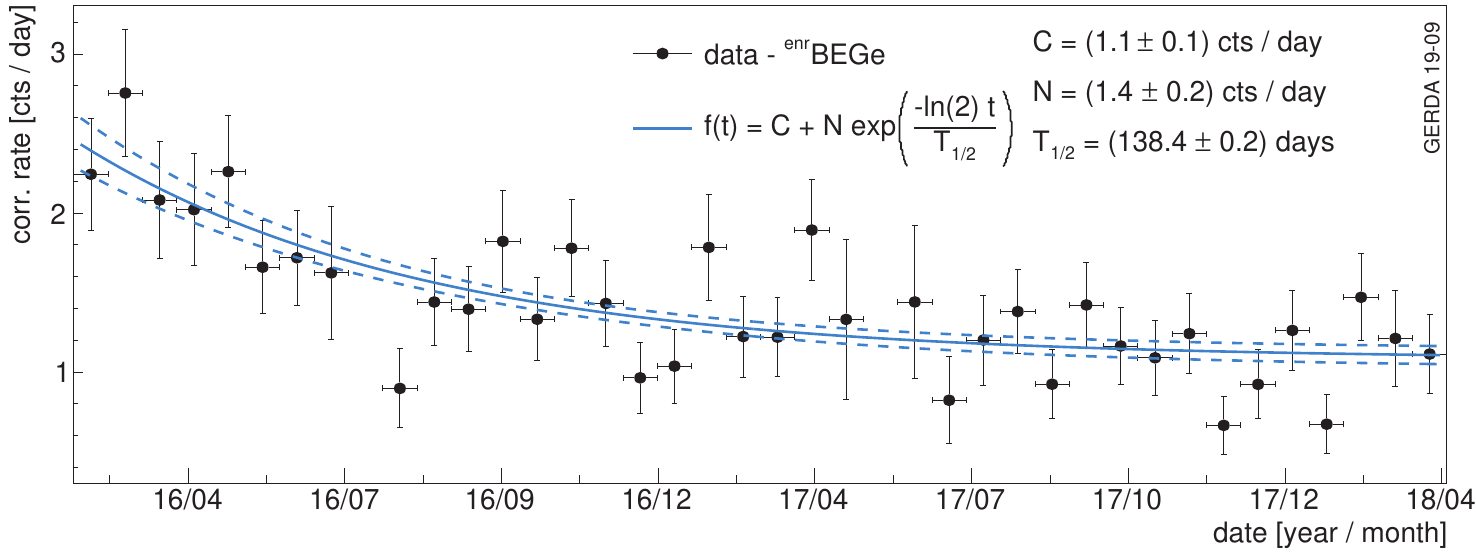}
  \includegraphics[width=0.8\textwidth]{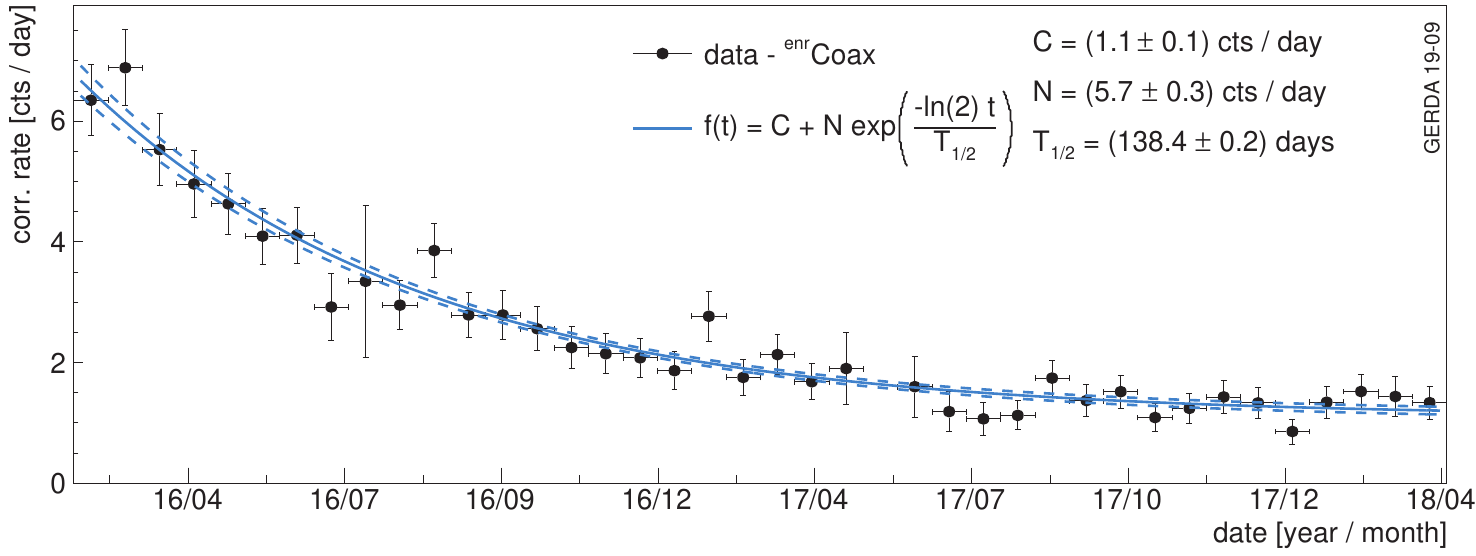}
  \caption{%
    $\alpha$-events time distribution in $[3500,5250]$~keV with a
    binning of 20~days for 27 $^\text{enr}$BEGe (top) and 7
    $^\text{enr}$Coax (bottom) detectors.
  }\label{fig:timealpha}
\end{figure}

The log-likelihood can be written as a sum:
\[
  \log \mathcal{L}_\alpha^\text{time}(C,N \,|\, n) =
  \sum_{i=1}^{N_\text{bins}} n_i \cdot
  \log\nu_i - \nu_i - \log n_i!
\]
We select only detectors that were ON or in anti-coincidence
mode\footnote{Detectors in anti-coincidence are not well
energy-calibrated and generally discarded in data analysis. Here, we are
not interested in the precise energy of an event because the selected
energy window is large with respect to a possible miscalibration.} in
the full data taking period. In this way we avoid bias due to selection
or deselection of particularly contaminated detectors. Furthermore, we
exclude the initial data-taking period between December 2015 to January
2016 from the following analysis because of detector instabilities after
the \phasetwo\ upgrade works. The analyzed data span from 25$^\text{th}$
January 2016 to 3$^\text{rd}$ April 2018 and are split into two data
sets according to detector type, containing 27 $^\text{enr}$BEGe and 7
$^\text{enr}$Coax detectors. The fit results are shown in
\autoref{fig:timealpha} and listed in \autoref{tab:timealpha}. For the
$^\text{enr}$BEGe data set we find that about half of the initial
contamination decays exponentially while for the $^\text{enr}$Coax data
set the ratio of $N$ to $C$ is about 5 to 1. After several \Po\
half-lives we expect a stable rate of $\sim1~\alpha$/day in either data
set.

\begin{table}[tbp]
  \centering
  \caption{%
    Results of the $\alpha$-events time distribution analysis in
    $[3500,5250]$~keV with a binning of 20~days for 27 $^\text{enr}$BEGe
    and 7 $^\text{enr}$Coax detectors.
  }\label{tab:timealpha}
  \begin{tabular}{cccccr@{ }l}
  \toprule
  \multirow{2}{*}{parameter} & \multirow{2}{*}{data} & \multirow{2}{*}{units} & \multirow{2}{*}{global mode} & \multicolumn{2}{c}{marg.~mode} \\
                             &                       &                        &                        &                              & \multicolumn{2}{c}{68\% C.I.} \\
  \midrule
  \multirow{2}{*}{$C$}  & $^\text{enr}$BEGe & \multirow{2}{*}{cts/day} & $1.06$  & 1.05 & $[1.00,1.12]$ \\
                        & $^\text{enr}$Coax &                          & $1.09$  & 1.09 & $[1.02,1.16]$ \\
  \multirow{2}{*}{$N$}  & $^\text{enr}$BEGe & \multirow{2}{*}{cts/day} & $1.32$ & 1.33 & $[1.13,1.53]$ \\
                        & $^\text{enr}$Coax &                          & $5.71$ & 5.70 & $[5.42,6.01]$ \\
  \bottomrule
\end{tabular}

\end{table}


\section{Monte Carlo simulations and probability density functions}%
\label{apdx:pdfs}

Background components that were identified in the energy spectra
(see~\autoref{subsec:data}) or in radio-purity screening
measurements~\cite{Agostini2018a} are simulated using the \mage\
software~\cite{boswell2011} based on
\geant~\cite{agostinelli2002,allison2006,Allison2016}.

The \gerdatwo\ detectors, their arrangement in seven strings as well as
the LAr instrumentation are implemented into \mage. A graphic rendering
of the relevant implemented hardware components is presented
in~\autoref{fig:magevolumes}. Simulations of radioactive contaminations
in the following hardware components are performed: in the bulk and on
the \pplus\ and \nplus\ surfaces of the germanium detectors, in the LAr,
detector holder bars and plates, nylon mini-shrouds, LAr veto system
(i.e.~the fiber shroud, SiPMs, copper shrouds and photomultipliers) and
in the signal and high-voltage flexible flat cables. The primary
spectrum of the two electrons emitted in the \twonu\ decay is sampled
according to the distribution given in reference~\cite{Tretyak1995}
implemented in \textsc{Decay0}~\cite{Ponkratenko2000}. Note that the
thickness of the detector assembly components are significantly smaller
than the mean free path of the relevant simulated $\gamma$-particles in
the given material, thus, no significant difference can be expected
between the resulting spectra of bulk and surface contaminations. The
detectors \nplus\ contact thicknesses are implemented according to the
values reported in references~\cite{Agostini2014,
GERDAcollaboration2019}.

The \kvz\ decays (except for surface contaminations) are simulated
homogeneously distributed in the relevant LAr volume. The following LAr
volumes are chosen for the background model: the first is a cylinder
centered on the detector array ($h=250$~cm, $r=100$~cm, simply referred
to as ``homogeneous'' or abbreviated to ``hom.''~in the following)
subsequently divided into the volume enclosed by the mini-shrouds and
the remaining one (outside the mini-shrouds); the second is a cylinder
($h=100$~cm, $r=25$~cm) positioned just above the array and the
remaining seven are smaller cylinders ($h=20$~cm, $r=5$~cm), each one
positioned just above each of the seven detector strings.

\begin{figure}[t]
  \centering
  \subfloat[%
    $\alpha$-decays from the \Ra\ decay sub-chain (\Ra, \Rn, $^{218}$Po
    and $^{214}$Po) on the detectors \pplus\ contact surface for
    different depths of the inactive contact layer. The isotope
    contamination is assumed to halve at each decay step, because of the
    recoil of the nuclei in LAr.\label{fig:pdfs:amodel:Ra}
  ]{\includegraphics[width=0.45\textwidth]{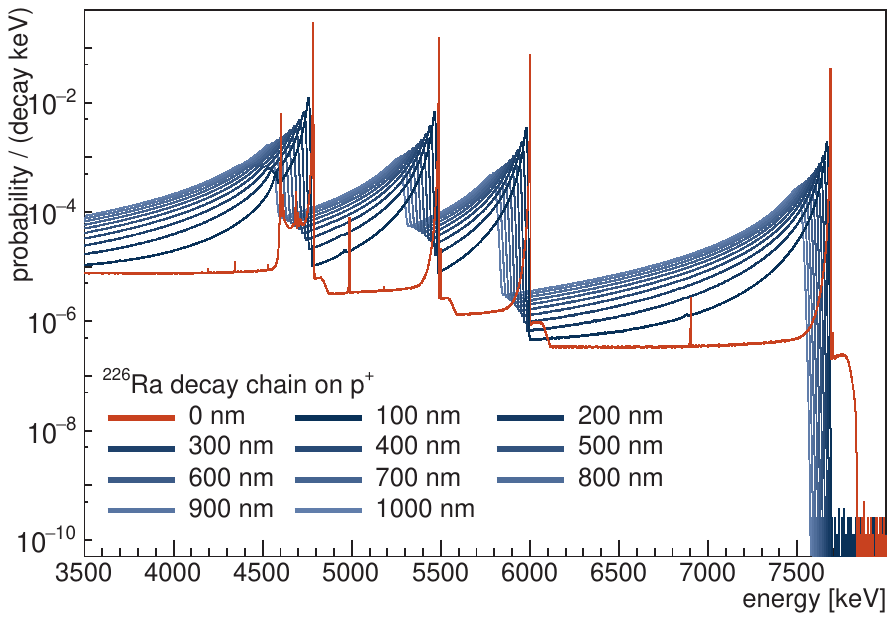}}
    \hspace{10pt}
  \subfloat[%
    \Pbh\ and \Bih\ (\Uh\ chain) contaminations far from (fiber-shroud)
    and close to (mini-shrouds) the detector array. A variable binning
    is adopted for visualization purposes.\label{fig:pdfs:gmodel:U}%
  ]{\includegraphics[width=0.45\textwidth]{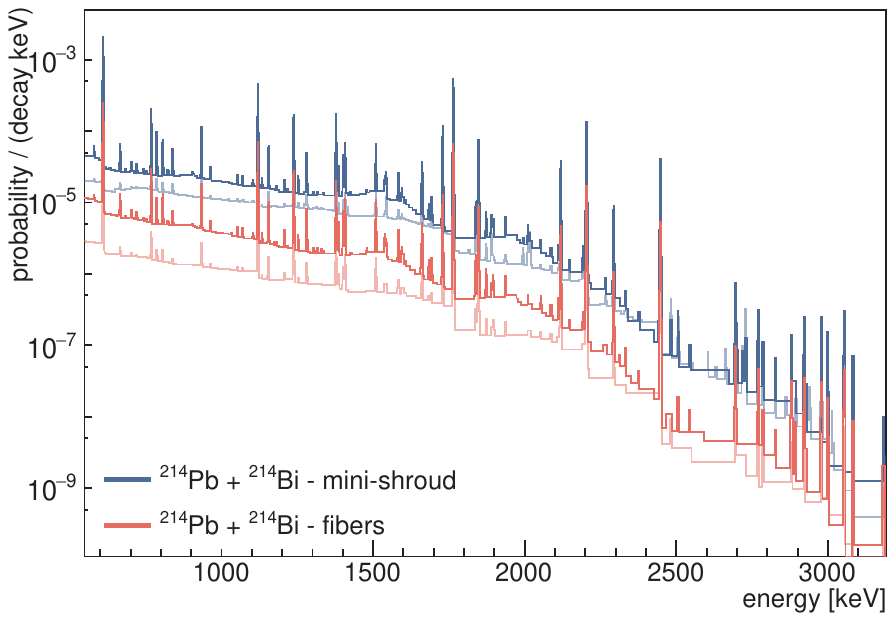}}
  \caption{%
    PDFs in the full energy domain. All PDFs are normalized to the
    number of simulated primary decays.
  }\label{fig:apdx:pdfs:gmodel}
\end{figure}

On top of the \mage\ simulations a post-processing step is performed to
compute the Probability Density Functions (PDFs) used to model the
\gerda\ data in the statistical analysis. This includes folding in
run-time dependent information, i.e.~the detector status in each physics
run, the finite energy resolution and threshold of each detector. All
PDFs presented in the following are computed using the run-time
parameters of the data sets described in~\autoref{subsec:data}. A
selection of the PDFs projected in energy space and normalized to the
number of simulated primary decays, are displayed
in~\autoref{fig:pdfs:gmodel} and~\autoref{fig:apdx:pdfs:gmodel}.

\begin{figure}[p]
  \centering
  \subfloat[%
    \kvn\ in different setup locations and \twonu\ in Ge,
    \Mokvn\ data set.\label{fig:apdx:pdfs:kmodel:K40}%
  ]{\includegraphics[width=0.48\textwidth]{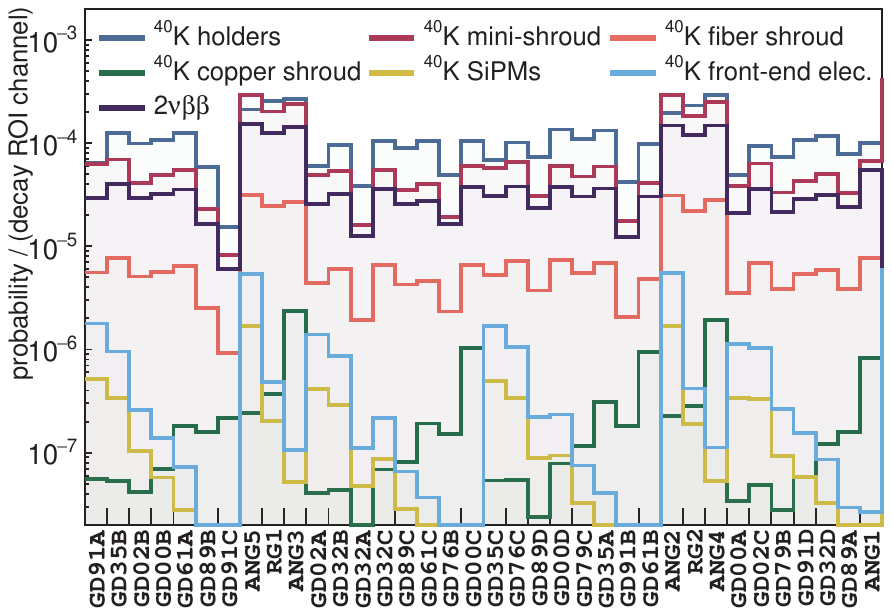}}
  \hspace{5pt}
  \subfloat[%
    \kvn\ located close to each single mini-shroud, \Mokvn\
    data set.\label{fig:apdx:pdfs:kmodel:K40sep}%
  ]{\includegraphics[width=0.48\textwidth]{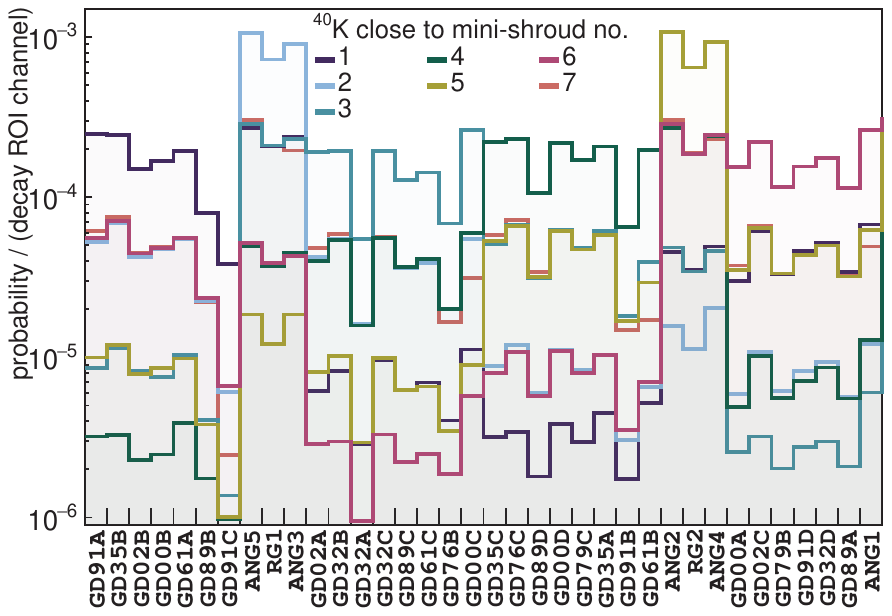}}
  \hspace{5pt}
  \subfloat[%
    \kvn\ in different setup locations, \Mtkvn\
    data set.\label{fig:apdx:pdfs:kmodel:M2K40}%
  ]{\includegraphics[width=0.48\textwidth]{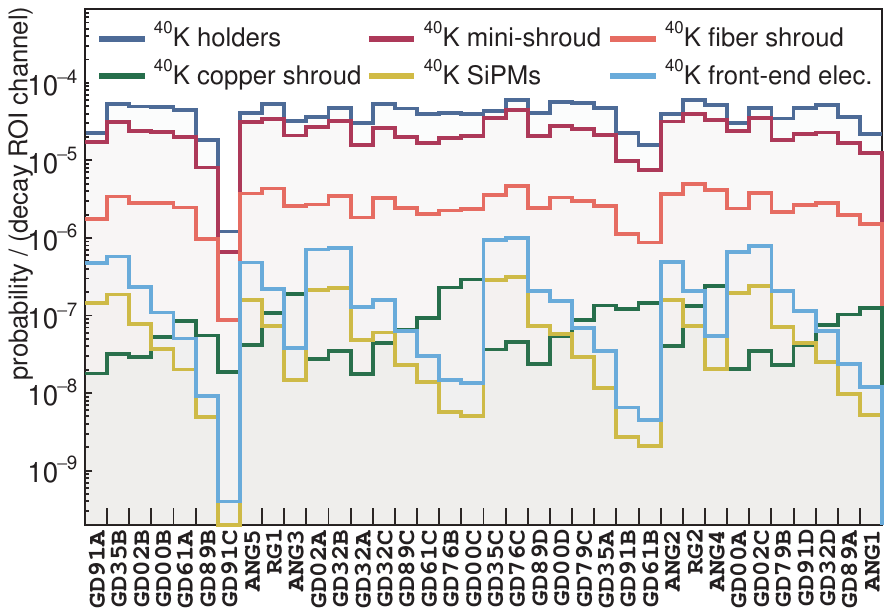}}
  \hspace{5pt}
  \subfloat[%
    \kvn\ located close to each single mini-shroud, \Mtkvn\
    data set.\label{fig:apdx:pdfs:kmodel:M2K40sep}%
  ]{\includegraphics[width=0.48\textwidth]{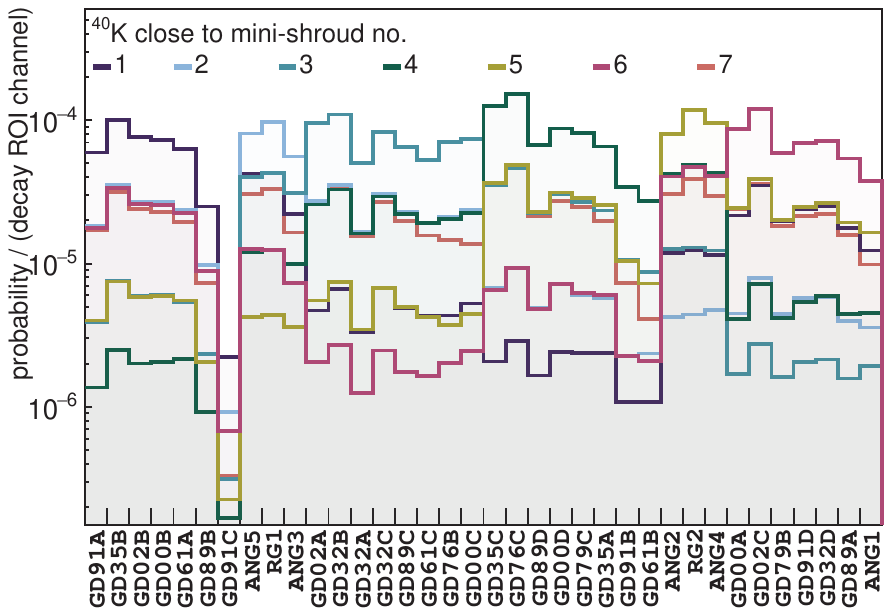}} \\
  \hspace{5pt}
  \subfloat[%
    \kvz\ in LAr above each single mini-shroud, \Mokvz\
    data set.\label{fig:apdx:pdfs:kmodel:K42sep}%
  ]{\includegraphics[width=0.48\textwidth]{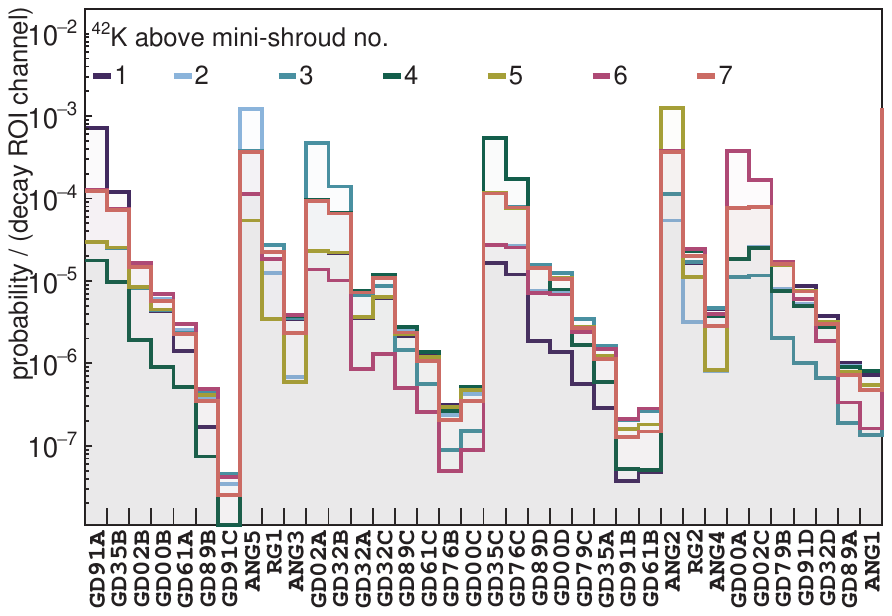}}
  \hspace{5pt}
  \subfloat[%
    \kvz\ in different setup locations, \Mtkvz\
    data set.\label{fig:apdx:pdfs:kmodel:M2K42}%
  ]{\includegraphics[width=0.48\textwidth]{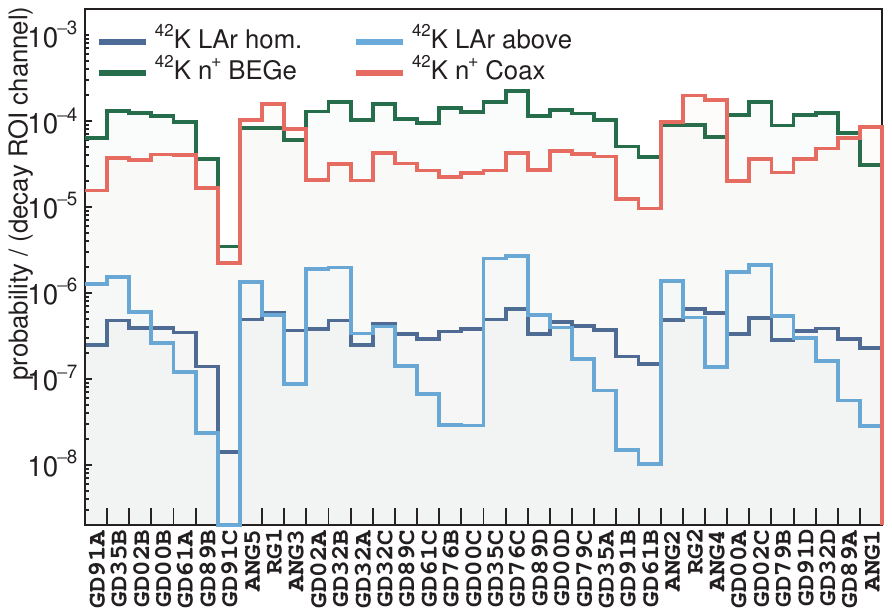}}
  \caption{%
    PDFs binned in detector space for the potassium tracking analysis. 
    All PDFs are normalized to the number of simulated primary decays.
  }\label{fig:apdx:pdfs:kmodel}
\end{figure}

For the potassium tracking analysis PDFs binned in detector space are
used to model the data. The rotationally symmetric single-detector PDFs
for the \kvn\ and \kvz\ energy windows are shown in
\autoref{fig:pdfs:kmodel:K42} and \autoref{fig:apdx:pdfs:kmodel:K40}.
For two-detector events the same representation style as
in~\autoref{fig:kmodel:spc} is used: projections of the two-dimensional
histograms on their axis are summed, such that each two-detector event
enters the final histogram twice, in the two bins associated to the
respective detectors. They can be found in
\autoref{fig:apdx:pdfs:kmodel} together with the single-detector PDFs of
the rotationally asymmetric components. 

Common features can be noticed across the multitude of histogram shapes.
The event rate in single-detector data is generally higher in coaxial
detectors, due to their larger mass compared to BEGe detectors ---
maximal correlation between event rate and detector-by-detector exposure
can be found in the \twonu\ PDF in~\autoref{fig:pdfs:kmodel:K42}. This
feature is generally lost in the two-detector data: the coaxial
detectors larger volume allows to stop more efficiently
$\gamma$-particles that would otherwise escape and eventually deposit
energy in a second detector. Other similarities between different PDFs
can be attributed to detectors live-times, like in the case of
\texttt{GD91C}, which was inactive for a large fraction of the
\phasetwo\ exposure and thus generally registers a low number of counts.
The effects of asymmetrically distributed background contaminations are
easily recognizable in the shape of the PDFs.  Impurities located above
the detector array are mostly seen by the upper most detectors in each
string as can be seen for \kvn\ in the front-end electronics
in~\autoref{fig:apdx:pdfs:kmodel:K40} and
in~\autoref{fig:apdx:pdfs:kmodel:M2K40} and for \kvz\ above each
mini-shroud (see~\autoref{fig:apdx:pdfs:kmodel:K42sep} and
\autoref{fig:apdx:pdfs:kmodel:M2K40sep}). Rotationally asymmetric
components are mostly evident in a single string, see for example \kvn\
in single mini-shrouds in \autoref{fig:apdx:pdfs:kmodel:K40sep} and
\autoref{fig:apdx:pdfs:kmodel:M2K40sep}.

All $\alpha$-decays in the \Ra\ to \Pbl\ sub-chain and from \Po\ are
simulated on the \pplus\ detector surface separately and for different
thicknesses of the \pplus\ electrode. The \Ra\ chain is simulated
together under the assumption that in each $\alpha$-decay half of the
contamination is lost due to the recoil of the nucleus into the LAr. The
resulting PDFs are displayed in~\autoref{fig:pdfs:amodel:Po} and
\autoref{fig:pdfs:amodel:Ra}. The spectra exhibit a peak like structure
with a pronounced low-energy tail.  The maximum is shifted with respect
to the full emission energy due to the thickness of the \pplus\ contact.
The low-energy tail is characteristic for $\alpha$-decays; the
$\alpha$-particle is susceptible to the change in the contact thickness
when penetrating the detector surface under an incident angle and loses
part of its energy before reaching the active detector volume.

\clearpage
\bibliography{gerda-bkg-paper-ph2}

\end{document}